\documentclass[prb,aps,twocolumn]{revtex4}
\usepackage{bm}
\usepackage{amsmath}
\usepackage{graphicx}
\makeatother
\begin{document}
%
\title{N\'{e}el Order and Electron Spectral Functions in the Two-Dimensional
                    Hubbard Model: \\ a Spin-Charge Rotating Frame Approach}

\author{T. A. Zaleski}
\affiliation{
Institute of Low Temperature and Structure Research,\\
Polish Academy of Sciences,\\
POB 1410, 50-950 Wroc\l aw 2, Poland}
\author{T. K. Kope\'{c}}
\affiliation{
Institute of Low Temperature and Structure Research,\\
Polish Academy of Sciences,\\
POB 1410, 50-950 Wroc\l aw 2, Poland}

 \begin{abstract}
Using recently developed quantum SU(2)$\times$ U(1) rotor approach,
that provides   a  self-consistent treatment of the antiferromagnetic state  
 we have performed electronic  spectral function calculations for
the Hubbard model on the square lattice. 
The collective variables for charge and spin  are isolated in the form of the
space-time fluctuating U(1) phase field and rotating spin quantization axis governed
by the SU(2) symmetry, respectively. As a result interacting  electrons  appear as composite 
objects consisting of bare fermions  with attached U(1) and SU(2) gauge fields.
This allows us to write the fermion Green's function in the space-time domain as 
the product  CP$^{1}$ propagator  resulting  from the SU(2) gauge fields, 
U(1) phase propagator and the pseudo-fermion correlation function.
 As a result the problem of calculating the spectral line shapes 
 now becomes one of performing the convolution of  spin,  
charge and pseudo-fermion Green's functions.
The collective spin and charge fluctuations are governed by the effective
 actions that are derived from the Hubbard model for any value of the Coulomb interaction.
The emergence of a sharp peak in the electron spectral function 
in the antiferromagnetic  state  indicates the  decay of the 
electron  into separate spin and charge carrying
particle excitations.
\end{abstract}
\pacs{71.10.Fd,71.10.-w,71.10.Pm}	

\maketitle
%
\section{Introduction}
%
Recent high-resolution angle-resolved photoemission
spectroscopy (ARPES) studies revealed a complicated
character of electronic structure and quasiparticle (QP)
spectra in copper oxide superconductors.\cite{arpes}
In the discussion of photoemission on solids, and in
particular on the correlated electron systems,  the most
powerful and commonly used approach is based on the
Green's--function formalism. In this context, the propagation
of a single electron in a many-body system is described
by the time-ordered one-electron Green's propagator.
A common approach in describing strong electron
correlations is based on consideration of the Hubbard
model.\cite {hubbard}  It allows to study a moderate correlation
limit observed experimentally in cuprates and more
consistently takes into account  subtleties of the  electronic structure,
in particular, a  spectral  weight transfer.\cite{meind}
In  qualitative sense the Hubbard model serves as the standard
model of correlated electron systems, and has the same conceptional importance
for interacting electrons as the Ising model for classical statistical
mechanics.  Here,  the low number of explicit  parameters provides the
ideal condition for a thorough test of the power and quality of analytical
and numerical methods. As a matter of fact
intensive studies on this  model  have  revealed subtlety of the
results  and controversies depending
on approaches, and  approximations.
The two-dimensional (2D) Hubbard model was studied in Ref. \onlinecite{zlatic}. 
The single particle self-energy was calculated  by perturbation theory
with the strength  of the local Coulomb interaction as the
expansion parameter to the second order. The main effect of correlations
is the transfer of the spectral weight to the high energies.
The 2D Hubbard model on the square lattice was also studied in the presence 
of lattice distortions in the adiabatic approximation.\cite{lamas}
In the absence of distortions the weight of the logarithmic
singularity which characterizes the free system is reduced.
Large $U$ values give rise to the typical two-peaks situation corresponding
to the infinite $U$ limit.
In other works, the 2D Hubbard model was considered with the 
quantum Monte Carlo method,\cite{monte1,monte2,monte3} recently also in the dynamical cluster approach.\cite{vidhya} 
To solve the cluster problem the Hirsch-Fye quantum Monte
Carlo method has been combined with the maximum entropy
method to calculate the real frequency spectra.
The effect of larger clusters and interactions was also explored
however, these results were restricted by computational
limitations including, especially, the minus sign problem.\cite{juillet1}
Method based on numerical simulations for finite clusters
precludes, however, to study
subtle features of QP spectra due to poor energy and
wave-vector resolutions in small size clusters.\cite{bulut}
Thus careful analyzes of finite-size effects are important in numerical studies, especially
for low-energy excitations. In the dynamical mean field theory
(DMFT) the self-energy is treated
in the single-site approximation which is unable to describe
wave-vector dependent phenomena.\cite{dmft}
To overcome the difficulties present in  DMFT, various types of the dynamical cluster
theory were developed.\cite{cluster} 
In these methods only a restricted wave-vector and energy
resolutions can be achieved, depending on the size of the
clusters, while the physical interpretation of the origin of
an anomalous electronic structure in numerical methods
is not straightforward.
In the Two-Particle Self-Consistent (TPSC) approach,\cite{vilk1,vilk2} that is based 
on enforcing sum rules and conservation laws, rather than on diagrammatic perturbative 
methods, a Luttinger-Ward functional is parametrized by two irreducible vertices that 
are local in space-time. This generates random phase approximation-like (RPA) equations 
for spin and charge fluctuations. The approach has the simple physical appeal of RPA but 
it satisfies key constraints that are always violated by RPA, namely the Mermin-Wagner 
theorem and the Pauli principle.
The inherent difficulty of dealing with 
Hamiltonians appropriate for strongly correlated electronic systems 
originates from the non-perturbative nature of the problem and  the presence 
of several competing physical mechanisms.
In the fermionic systems with spin and
charge excitations, however, the situation is much more
complicated because dynamic quantum fluctuations are
important even in large dimensions. Moreover,
 short-range spatial correlations play a key role, in particular
magnetic correlations, leading to a strong tendency towards the
formation of singlet bonds, as well as pair correlations. These correlations deeply
affect the nature of quasiparticles.
A new approach
may therefore be needed, and a necessary requirement
for a theory aiming to capture the essential
physics of electrons correlated systems might be the inclusion
of the fundamental ingredients of the physics as the  underlying spin and charge symmetries
as well as the associated ordered states.
In this context the spectral properties of the  two dimensional 
systems with strong magnetic fluctuations were investigated
within the spin-fermion model\cite{chubukov}   in the quasistatic approach that
neglects the effect of dynamic spin fluctuations. The latter
approach yields the two-peak structure of the spectral
function for the antiferromagnetic correlations.

The other completely different approaches rest on
gauge theories that arise in  models of strongly
interacting electrons: it is often convenient to
change variables from electron operators to other
degrees of freedom that represents the intrinsic symmetries of the system
under study. This route turns out to be beneficial since
several quantum phases of matter (for instance
a antiferromagnet or a superconductor) may be characterized in
terms of the symmetries that are broken spontaneously
in that state. \cite{borejsza1, dupuis1}
We employ this idea
in the present paper to perform the one-particle
spectral function  calculations for
the  Hubbard model on the square lattice
 using  a recently developed the quantum SU(2)$\times$U(1) rotor approach.\cite{zal}
The collective variables for charge and spin  are isolated in the form of the
space--time fluctuating U(1) phase field and rotating spin quantization axis governed
by the SU(2) symmetry, respectively. As a result interacting  electrons  appear as a composite 
objects consisting of bare fermions  with attached U(1) and SU(2) gauge fields.
This allows us to write the fermion Green's function in the 
space-time domain as the product of the complex-projective (CP$^{1}$) propagator
(which results from the SU(2) gauge fields), U(1) phase propagator 
and the pseudo-fermion correlation function.
The problem of calculating the spectral line shapes  now becomes
 one of calculating the convolution of  spin,  charge and pseudo-fermion Green's functions.
The collective spin and charge fluctuations are governed by 
the effective actions that are derived from
the Hubbard model for any value of the Coulomb interaction.
We show that the method is useful for explicit
calculation of spectral properties, enabling 
systematic inclusion of fluctuations of the charge as well as
of the variable spin quantization axis.
We demonstrate  that the emergence of a sharp
peak in the electron spectral function in the antiferromagnetic  state points to the
electron decaying into separate spin and charge carrying
particle excitations.

The paper is organized as follows. After
introduction of the  model Hamiltonian in Section II, we present in Sec. III 
the transformations to the phase and spin angular variables
that reflect the basic symmetries of the Hubbard model.
Section IV is devoted to the derivation of the effective actions that govern the
behavior of the system in the charge, spin and pseudo-fermion sectors, respectively.
The self-consistent equations for the effective parameters of the model that
follows from the effective actions are summarized in Section V.
Section VI  is devoted to  a detailed analysis of
the electronic spectral functions.
Conclusions and discussions are given in Section VII.
A number of technical details that pertain to the derivation of
spectral functions is relegated to the Appendices.
%
\section{The Model}
%
A prototype of theoretical understanding of
the physics of  strongly correlated systems is
achieved by using simplified lattice fermionic systems, in
particular, the  Hubbard model given by the
Hamiltonian ${\cal H}\equiv{\cal H}_{t}+{\cal H}_{U}$:
\begin{eqnarray}
{\cal H}=-t\sum_{\langle{\bf r}{\bf r}'\rangle,\alpha}
[c_{\alpha}^{\dagger}({\bf r})c_{\alpha}({\bf r}')+
{\rm h.c.}]-\mu\sum_{{\bf r}}n({\bf r})+\mathcal{H}_{U},\label{mainham}
\end{eqnarray}
where the Hubbard interaction term is given by
\begin{equation}
\mathcal{H}_{U}=U\sum_{{\bf r}}n_{\uparrow}({\bf r})n_{\downarrow}({\bf r})
\end{equation}
and $n({\bf r})=n_{\uparrow}({\bf r})+n_{\downarrow}({\bf r})$ is
the number operator. Here, $\langle{\bf r},{\bf r}'\rangle$ runs
over the nearest-neighbor (n.n.) sites, $t$ is the hopping amplitude,
$U$ stands for the Coulomb repulsion, while the operator $c_{\alpha}^{\dagger}({\bf r})$
creates an electron with spin $\alpha=\uparrow(\equiv 1),\downarrow(\equiv 2)$ at the
lattice site ${\bf r}$, where $n_{\alpha}({\bf r})=c_{\alpha}^{\dagger}({\bf r})c_{\alpha}({\bf r})$.
The chemical potential $\mu$ controls the average number of electrons.
The kinetic energy operator ${\cal H}_{t}$ reflects
the electrons' itinerant features  while the interaction 
operator  ${\cal H}_{U}$ forces the electrons' correlated motion or even
their localization.

Since the partition function often serves as a starting point for the calculation
of thermodynamic properties, it is instructive to take a closer look at how this quantity
may be obtained within the path integral formalism. To this end
it is customary to introduce Grassmann fields, $c_{\alpha}({\bf r}\tau)$
depending on the {}``imaginary time\char`\"{} $0\le\tau\le\beta\equiv1/k_{B}T$,
(with $T$ being the temperature) that satisfy the anti--periodic
condition $c_{\alpha}({\bf r}\tau)=-c_{\alpha}({\bf r}\tau+\beta)$,
to write the path integral for the statistical sum 
\begin{equation}
{\cal Z}=\int\left[{\cal D}\bar{c}{\cal D}c\right]e^{-{\cal S}[\bar{c},c]}
\end{equation}
with the fermionic action 
\begin{eqnarray}
{\cal S}[\bar{c},c]={\cal S}_{B}[\bar{c},c]+\int_{0}^{\beta}d\tau{\cal H}[\bar{c},c],
\end{eqnarray}
 which contains the fermionic Berry term\cite{berry} 
\begin{equation}
{\cal S}_{B}[\bar{c},c]=\sum_{{\bf r}\alpha}\int_{0}^{\beta}
d\tau\bar{c}_{\alpha}({\bf r}\tau)\partial_{\tau}c_{\alpha}({\bf r}\tau)
\end{equation}
 that will play an important role in our considerations.
%
\section{Spin-charge rotating reference frame}
%
The  spin-rotational symmetry present in the Hubbard Hamiltonian is
instrumental for obtaining proper low energetic properties. Therefore,  it is crucial
to construct a  theoretical formulation that  naturally preserves
this symmetry. In particular, one should consider the spin-quantization
axis to be a priori arbitrary and integrate over all possible directions
in the partition function. It can be achieved when the density--density product
in Eq.(\ref{mainham}) is written, following Ref. \onlinecite{schulz}, in
a spin-rotational invariant way:
 \begin{equation}
{\cal H}_{U}=U\sum_{{\bf r}}\left\{ \frac{1}{4}n^{2}({\bf r}\tau)
-\left[{\bf \Omega}({\bf r}\tau)\cdot{\bf S}({\bf r}\tau)\right]^{2}\right\} ,\label{huu}
\end{equation}
 where $S^{a}({\bf r}\tau)=\frac{1}{2}\sum_{\alpha\alpha'}
c_{\alpha}^{\dagger}({\bf r}\tau)\hat{\sigma}_{\alpha\alpha'}^{a}c_{\alpha'}({\bf r}\tau)$
denotes the vector spin operator ($a=x,y,z$) with $\hat{\sigma}^{a}$
being the Pauli matrices. The unit vector 
\begin{eqnarray}
{\bf \Omega}({\bf r}\tau) & = & [\sin\vartheta({\bf r}\tau)\cos\varphi({\bf r}\tau),
\sin\vartheta({\bf r}\tau)\sin\varphi({\bf r}\tau),\nonumber \\
 &  & \cos\vartheta({\bf r}\tau)]
\end{eqnarray}
written in terms of polar angles labels varying in space-time spin
quantization axis. The explicit spin--rotation invariance comes from 
the angular integration over ${\bf \Omega}({\bf r}\tau)$
at each site and time:
\begin{equation}
{\cal Z}=\int[{\cal D}{\bf \Omega}]\int\left[{\cal D}\bar{c}{\cal D}c\right]
e^{-{\cal S}[{\bf \Omega},\bar{c},c]},
\end{equation}
where $[{\cal D}{\bf \Omega}]\equiv\prod_{{\bf r}\tau_{k}}
\frac{\sin\vartheta({\bf r}\tau_{k})d\vartheta({\bf r}\tau_{k})d\varphi({\bf r}\tau_{k})}{4\pi}$
is the spin-angular integration measure. 
%
\subsection{Hubbard-Stratonovich decoupling}
%
The spin and charge density terms of the Hamiltonian in Eq.(\ref{huu}) are of 
the fourth order in fermionic operators, so they must be decoupled using
Hubbard-Stratonovich (HS) formula\cite{hs} with the
auxiliary fields $\varrho({\bf r}\tau)$ and $iV({\bf r}\tau)$ respectively.
The partition function can be written in the form\cite{popov} 
\begin{eqnarray}
{\cal Z} & = & \int[{\cal D}{\bf \Omega}]\int[{\cal D}V{\cal D}
\varrho]\int\left[{\cal D}\bar{c}{\cal D}c\right]e^{-{\cal S}
\left[{\bf \Omega},V,\varrho,\bar{c},c\right]}.\label{zfun}
\end{eqnarray}
Consequently, the effective action reads:
\begin{eqnarray}
{\cal S}\left[{\bf \Omega},V,\varrho,\bar{c},c\right] & = 
& \sum_{{\bf r}}\int_{0}^{\beta}d\tau\left[\frac{\varrho^{2}
({\bf r}\tau)}{U}+\frac{V^{2}({\bf r}\tau)}{U}\right.\nonumber \\
 & + & \left.iV({\bf r}\tau)n({\bf r}\tau)+2\varrho({\bf r}\tau)
{\bf \Omega}({\bf r}\tau)\cdot{\bf S}({\bf r}\tau)\right]\nonumber \\
 & + & {\cal S}_{B}[\bar{c},c]+\int_{0}^{\beta}d\tau{\cal H}_{t}[\bar{c},c].\label{sa}
\end{eqnarray}
Since, $U$ is the largest energy in the problem, the simple Hartree--Fock 
theory won't work. To proceed, one has to isolate strongly
fluctuating modes generated by the Hubbard term according to the charge
U(1) and spin SU(2) symmetries. 
%
\subsection{U(1) charge frame}
%
Now, we switch
from the particle-number representation to the conjugate
phase representation of the electronic degrees of freedom that is governed by the compact U(1)
group. To this end the second quantized
Hamiltonian of the model is translated to the phase
representation with the help of the topologically constrained
path integral formalism.\cite{schulman} As a result the electrons emerge as
composite particles consisting of of spin-carrying neutral fermions
and topological charged bosons in a form of a "flux
tubes" with the quantum phase variable dual to the local
electron density.
To this end we write the fluctuating {}``imaginary chemical potential\char`\"{}
$iV({\bf r}\tau)$ as a sum of a static $V_{0}({\bf r})$ and periodic
function 
\begin{equation}
V({\bf r}\tau)=V_{0}({\bf r})+\tilde{V}({\bf r}\tau),
\end{equation}
where, using Fourier series 
\begin{equation}
\tilde{V}({\bf r}\tau)=\frac{1}{\beta}\sum_{n=1}^{\infty}
[\tilde{V}({\bf r}\omega_{n})e^{i\omega_{n}\tau}+c.c.]
\end{equation}
with $\omega_{n}=2\pi n/\beta$ ($n=0,\pm1,\pm2$) being the (Bose)
Matsubara frequencies. Now, we introduce the U(1) \textit{phase}  field
$\phi({\bf r}\tau)$ via the Faraday--type relation\cite{kopec}
\begin{equation}
\dot{\phi}({\bf r}\tau)\equiv\frac{\partial\phi({\bf r}\tau)}
{\partial\tau}=e^{-i\phi({\bf r}\tau)}\frac{1}{i}\frac{\partial}
{\partial\tau}e^{i\phi({\bf r}\tau)}=\tilde{V}({\bf r}\tau).\label{jos}
\end{equation}
 Furthermore, by performing the local gauge transformation to the
{new} fermionic variables $f_{\alpha}({\bf r}\tau)$: 
\begin{eqnarray}
\left[\begin{array}{c}
c_{\alpha}({\bf r}\tau)\\
\bar{c}_{\alpha}({\bf r}\tau)\end{array}\right]=\left[\begin{array}{cc}
z({\bf r}\tau) & 0\\
0 & \bar{z}({\bf r}\tau)\end{array}\right]\left[\begin{array}{c}
f_{\alpha}({\bf r}\tau)\\
\bar{f}_{\alpha}({\bf r}\tau)\end{array}\right],\label{sing1}
\end{eqnarray}
 where the unimodular parameter $|z({\bf r}\tau)|^{2}=1$ satisfies
$z({\bf r}\tau)=e^{i\phi({\bf r}\tau)}$, we remove the imaginary
term $i\int_{0}^{\beta}d\tau\tilde{V}({\bf r}\tau)n({\bf r}\tau)$
for all the Fourier modes of the $V({\bf r}\tau)$ field, except for
the zero frequency. 
We point out here that a similar phase representation
was developed  in the context of Coulomb blockade
in mesoscopic systems.\cite{schon}
%
\subsection{SU(2) spin frame}
%
Subsequent SU(2) transformation from $f_{\alpha}({\bf r}\tau)$ to
$h_{\alpha}({\bf r}\tau)$ operators, 
\begin{eqnarray}
\left[f_{\uparrow}({\bf r}\tau),\,\,\, f_{\downarrow}({\bf r}\tau)\right]
 & = & {\bf R}({\bf r}\tau)\left[\begin{array}{c}
h_{\uparrow}({\bf r}\tau)\\
h_{\downarrow}({\bf r}\tau)\end{array}\right],\nonumber \\
{\bf R}({\bf r}\tau) & \equiv & \left[\begin{array}{cc}
\zeta_{\uparrow}({\bf r}\tau) & -\bar{\zeta}_{\downarrow}({\bf r}\tau)\\
\zeta_{\downarrow}({\bf r}\tau) & \bar{\zeta}_{\uparrow}({\bf r}
\tau)\end{array}\right]\label{sing2}
\end{eqnarray}
 with the constraint 
\begin{equation}
|\zeta_{\uparrow}({\bf r}\tau)|^{2}+|\zeta_{\downarrow}({\bf r}\tau)
|^{2}=1\label{eq:CP1constraint}
\end{equation}
takes away the rotational dependence on ${\bf \Omega}({\bf r}\tau)$
in the spin sector. This is done by means of the Hopf map \cite{fradkin}
\begin{equation}
{\bf R}({\bf r}\tau)\hat{\sigma}^{z}{\bf R}^{\dagger}({\bf r}\tau)
=\hat{\sigma}\cdot{\bf \Omega}({\bf r}\tau)
\end{equation}
that is based on the enlargement from two-sphere $S_{2}$ to the three-sphere
$S_{3}\sim SU(2)$. The unimodular constraint in Eq.(\ref{eq:CP1constraint}) 
can be resolved by using
the parametrization 
\begin{eqnarray}
\zeta_{1\uparrow}({\bf r}\tau) & = & e^{-\frac{i}{2}[\varphi({\bf r}\tau)
+\chi({\bf r}\tau)]}\cos\left[\frac{\vartheta({\bf r}\tau)}{2}\right]\nonumber \\
\zeta_{\downarrow}({\bf r}\tau) & = & e^{\frac{i}{2}[\varphi({\bf r}\tau)
-\chi({\bf r}\tau)]}\sin\left[\frac{\vartheta({\bf r}\tau)}{2}\right]\label{cp1}
\end{eqnarray}
 with the Euler angular variables $\varphi({\bf r}\tau),\,\vartheta({\bf r}\tau)$
and $\chi({\bf r}\tau)$, respectively. Here, the extra variable $\chi({\bf r}\tau)$
represents the U(1) gauge freedom of the theory as a consequence of
$S_{2}\to S_{3}$ mapping. One can summarize Eqs (\ref{sing1}) and
(\ref{sing2}) by the single joint gauge transformation exhibiting
electron operator factorization 
\begin{eqnarray}
c_{\alpha}({\bf r}\tau) & = & \sum_{\alpha'}
\mathcal{U}_{\alpha\alpha'}({\bf r}\tau)h_{\alpha'}({\bf r}\tau),\nonumber \\
\mathcal{U}({\bf r}\tau) & = & z({\bf r}\tau){\bf R}({\bf r}\tau),
\end{eqnarray}
where $\mathcal{U}\left(\mathbf{r}\tau\right)$ is a U(2) matrix which
rotates the charge and spin-quantization axis at site ${\bf r}$ and time $\tau$.
This reflects the composite nature of the interacting electron formed
from bosonic spin and charge degrees of freedom given by 
${R}_{\alpha\alpha'}({\bf r}\tau)$
and $z({\bf r}\tau)$, respectively as well as remaining fermionic
part $h_{\alpha}({\bf r}\tau)$. Accordingly, the integration measure
over the group manifold becomes 
\begin{eqnarray}
 &  & \int[{\cal D}\phi{\cal D}{\bf \Omega}]\equiv\sum_{\{m({\bf r})\}}
\prod_{{\bf r}}\int_{0}^{2\pi}d\phi_{0}({\bf r})\int d{\bf \Omega}_{0}({\bf r})\nonumber \\
 &  & \times\int\limits _{{\bf \Omega}({\bf r}0)={\bf \Omega}_{0}}^{{\bf \Omega}
({\bf r}\beta)={\bf \Omega}_{0}}{\cal D}{\bf \Omega}
({\bf r}\tau)\int\limits _{\phi({\bf r}0)=
\phi_{0}({\bf r})}^{\phi({\bf r}\beta)=
\phi_{0}({\bf r})+2\pi m({\bf r})}{\cal D}\phi({\bf r}\tau),\label{measure}
\end{eqnarray}
 where $\int d{\bf \Omega}\dots=\frac{1}{4\pi}\int_{0}^{\pi}\sin\theta d\vartheta\int_{0}^{2\pi}d\varphi\dots$
and $[{\cal D}{\bf \Omega}({\bf r}\tau)]=\prod_{k}d{\bf \Omega}({\bf r}\tau_{k})$.
Here, $m\in{Z}$ labels equivalence classes of homotopically connected
paths \cite{schulman} for the U(1) group. 
%
\subsection{Solutions for $V_{0}({\bf r})$ and $\varrho({\bf r}\tau)$ }
%
Once can anticipate that spatial and temporal fluctuations of the fields
$V_{0}({\bf r})$ and $\varrho({\bf r}\tau)$ will be energetically
penalized, since they are linked to the high energy scale set by $U$
and decouple from the angular and
phase variables. Therefore, in order to make further progress 
we next subject the corresponding functionals to a saddle point analysis. The expectation
value of the static (zero frequency) part of the fluctuating 
potential $V_{0}(\mathbf{r})$ calculated by the saddle point method
to give
 \begin{equation}
V_{0}(\mathbf{r})=i\left(\mu-\frac{U}{2}n_{h}\right)\equiv i\bar{\mu},
\end{equation}
 where $\bar{\mu}$ is the chemical potential with a Hartree shift
originating from the saddle-point value of the static variable $V_{0}({\bf r})$
with $n_{h}=n_{h\uparrow}+n_{h\downarrow}$ and
 $n_{h\alpha}=\langle\bar{h}_{\alpha}({\bf r}\tau)h_{\alpha}({\bf r}\tau)\rangle$.
Similarly in the magnetic sector we have
\begin{eqnarray}
\rho({\bf r}\tau)=
(-1)^{{\bf r}}\Delta_{c},\label{spaff}
\end{eqnarray}
 where $\Delta_{c}=U\langle S^{z}({\bf r}\tau\rangle$ sets the magnitude
for the Mott-charge gap. The solution delineated in Eq.(\ref{spaff})
correspond to the saddle point of the antifferomagnetic type
(with staggering $\Delta_{c}$) .
Note that the notion ``antifferomagnetic" here does not
mean an actual long--range ordering - for this the angular spin-quantization
variables ${\bf \Omega}({\bf r}\tau)$ have to be ordered as well. 
The mean-field parameter $\Delta_c$
has to be determined by the stationary points of the action,
e.g., by the mean-field equations that will be derived more
explicitly later on for our special purpose.
%
\section{Effective actions}
%
\subsection{Total fermionic phase-angular action}
%
In the new variables the action in Eq.(\ref{sa}) assumes the form
\begin{eqnarray}
 &  & {\cal S}\left[{\bf \Omega},\phi,\varrho,\bar{h},h\right]
={\cal S}_{B}[\bar{h},h]+\int_{0}^{\beta}d\tau{\cal H}_{{\bf \Omega,\phi}}[\rho,\bar{h},h]
\nonumber \\
 &  & \,\,\,\,\,+{\cal S}_{0}\left[\phi\right]+2\sum_{{\bf r}}
\int_{0}^{\beta}d\tau{\bf A}
({\bf r}\tau)\cdot{\bf S}_{h}({\bf r}\tau),
\end{eqnarray}
where
 \begin{equation}
{\bf S}_{h}({\bf r}\tau)=\frac{1}{2}\sum_{\alpha\gamma}\bar{h}_{\alpha}({\bf r}
\tau)\hat{\sigma}_{\alpha\gamma}h_{\gamma}({\bf r}\tau).
\end{equation}
Furthermore, 
\begin{eqnarray}
S_{0}[\phi]=\sum_{{\bf r}}\int_{0}^{\beta}d\tau\left[\frac{\dot{\phi}^{2}
({\bf r}\tau)}{U}+\frac{1}{i}\frac{2\mu}{U}\dot{\phi}({\bf r}\tau)\right]\label{sphi}
\end{eqnarray}
stands for the kinetic and Berry term of the U(1) phase field in
the charge sector. The SU(2) gauge transformation in Eq.(\ref{sing2})
and the fermionic Berry term generate SU(2) potentials given by 
\begin{eqnarray}
{\bf R}^{\dagger}({\bf r}\tau)\partial_{\tau}{\bf R}({\bf r}\tau)
 & = & {\bf R}^{\dagger}\left(\dot{\varphi}\frac{\partial}{\partial\varphi}
+\dot{\vartheta}\frac{\partial}{\partial\vartheta}+\dot{\chi}\frac{\partial}
{\partial\chi}\right){\bf R}\nonumber \\
 & = & -\bm{\sigma}\cdot{\bf A}({\bf r}\tau),
\end{eqnarray}
where 
\begin{eqnarray}
A^{x}({\bf r}\tau) & = & \frac{i}{2}\dot{\vartheta}({\bf r}\tau)\sin\chi({\bf r}\tau)
-\frac{i}{2}\dot{\varphi}({\bf r}\tau)\sin\theta({\bf r}\tau)\cos\chi({\bf r}\tau)
\nonumber \\
A^{y}({\bf r}\tau) & = & \frac{i}{2}\dot{\vartheta}({\bf r}\tau)\cos\chi({\bf r}\tau)
+\frac{i}{2}\dot{\varphi}({\bf r}\tau)\sin\theta({\bf r}\tau)\sin\chi({\bf r}\tau)
\nonumber \\
A^{z}({\bf r}\tau) & = & \frac{i}{2}\dot{\varphi}({\bf r}\tau)\cos\vartheta({\bf r}\tau)
+\frac{i}{2}\dot{\chi}({\bf r}\tau).
\end{eqnarray}
are the SU(2) gauge potentials. The fermionic sector, in turn, is
governed by the effective Hamiltonian
\begin{equation}
\mathcal{H}_{\Omega,\phi}=\mathcal{H}_{\Omega,\phi}^{(\rho)}+\mathcal{H}_{\Omega,\phi}^{(t)},
\end{equation}
where
\begin{eqnarray}
{\cal H}_{{\bf \Omega,\phi}}^{(\rho)} & = & 
\sum_{{\bf r}}(-1)^{{\bf r}}\Delta_{c}[\bar{h}_{\uparrow}({\bf r}\tau)h_{\uparrow}
({\bf r}\tau)-\bar{h}_{\downarrow}({\bf r}\tau)h_{\downarrow}({\bf r}\tau)]
\nonumber \\
{\cal H}_{{\bf \Omega,\phi}}^{(t)} & = & -t\sum_{\langle{\bf r},{\bf r}'\rangle}
\sum_{\alpha\gamma}\left[\mathcal{U}^{\dagger}({\bf r}\tau)
\mathcal{U}({\bf r'}\tau)\right]_{\alpha\gamma}\bar{h}_{\alpha}({\bf r}\tau)h_{\gamma}({\bf r}'\tau)
\nonumber \\
 & - & \bar{\mu}\sum_{{\bf r}\alpha}\bar{h}_{\alpha}({\bf r}\tau)h_{\alpha}({\bf r}\tau).
\label{explicit}
\end{eqnarray}
 The result of the gauge transformations is that we have managed to
cast the strongly correlated problem into a system of mutually non-interacting
pseudo fermions, submerged in the bath of strongly fluctuating U(1) and SU(2)
fields whose dynamics is governed by the energy scale set by the Coulomb
interaction $U$ coupled to fermions via hopping term and with the
Zeeman-type contribution with the massive field ${\varrho}({\bf r}\tau)$.
%
\subsection{Charge (phase) action}
%
\begin{figure}
\includegraphics[scale=0.7]{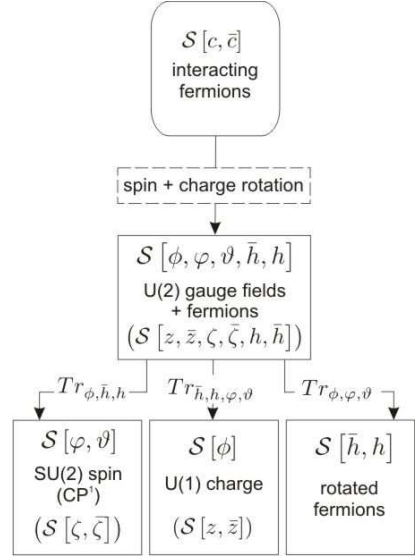}
\caption{Scheme for the derivation  of effective actions. The initial purely fermionic action
of the Hubbard model  is transformed into   effective actions in spin,
charge and fermionic sectors by performing the trace over selected set of variables. }
\end{figure}
%
In systems with  Coulomb interactions,
the phase variable dual to the  charge is an important
collective field. 
We start with a partition function for charge sector
\begin{eqnarray}
{\cal Z} & = & \int\left[{\cal D}\phi\right]e^{-\mathcal{S}[\phi]},
\end{eqnarray}
where the charge action requires tracing over fermionic and angular
SU(2) variables
\begin{eqnarray}
\mathcal{S}[\phi] & = & -\ln\int[{\cal D}\bar{h}{\cal D}h{\cal D}{\bf \Omega}]
e^{-{\cal S}[\varphi,\phi,\vartheta,\bar{h},h]}.
\end{eqnarray}
To proceed, it is convenient to replace the phase degrees of freedom
by the complex field 
\begin{eqnarray}
z\left(\mathbf{r}\tau\right) & = & e^{i\phi\left(\mathbf{r}\tau\right)}\nonumber \\
\bar{z}\left(\mathbf{r}\tau\right) & = & e^{-i\phi\left(\mathbf{r}\tau\right)},
\end{eqnarray}
which satisfies the periodic boundary condition $z\left(\mathbf{r}\beta\right)=z\left(\mathbf{r}0\right)$.
This can be done by implemented the Fadeev-Popov method with the Dirac
delta functional resolution of the unity:\cite{kopec1} 
\begin{eqnarray}
 &  & 1\equiv\int\left[\mathcal{D}\bar{z}\mathcal{D}z\right]\delta\left[\frac{1}{N}
\sum_{i}\left|z\left(\mathbf{r}\tau\right)\right|^{2}-1\right]\nonumber \\
 &  & \times\prod_{i}\delta\left[z\left(\mathbf{r}\tau\right)-e^{i\phi\left(\mathbf{r}
\tau\right)}\right]\delta\left[\bar{z}_{i}\left(\mathbf{r}\tau\right)
-e^{-i\phi\left(\mathbf{r}\tau\right)}\right],\label{popov}
\end{eqnarray}
 where we take $z\left(\mathbf{r}\tau\right)$ as continuous variable
but constrained (on the average) to have the unimodular value. We
can solve the constraint by introducing the Lagrange multiplier $\lambda$
which adds the quadratic terms (in the $z\left(\mathbf{r}\tau\right)$
fields) to the effective action. The partition function can be written
in form 
\begin{eqnarray}
\mathcal{Z} & = & \int_{-i\infty}^{+i\infty}\left[\frac{{\cal D}\lambda_{z}}{2\pi i}\right]
e^{-N\beta\mathcal{F}\left(\lambda_{z}\right)},
\end{eqnarray}
 where the free energy per site $\mathcal{F}=-\ln\mathcal{Z}/\beta N$
is given by:
 \begin{eqnarray}
\mathcal{F} & = & -\lambda_{z}-\frac{1}{N\beta}\ln\int\left[\mathcal{D}
\bar{z}\mathcal{D}z\right]e^{-{\cal S}_{\mathrm{eff}}[\bar{z},z]}\nonumber \\
{\cal S}_{\mathrm{eff}}[\bar{z},z] & = & \sum_{\langle\mathbf{r}\mathbf{r}'\rangle}
\int_{0}^{\beta}d\tau d\tau^{'}\left[\lambda_{z}\delta_{\mathbf{r}\mathbf{r}'}
\delta\left(\tau-\tau'\right)\right]\nonumber \\
 &  & \left.-\mathcal{\gamma}\left(\mathbf{r}\tau,\mathbf{r}'\tau'\right)\right]
\bar{z}(\mathbf{r}\tau)z(\mathbf{r}'\tau'),\label{free energy}
\end{eqnarray}
 where 
\begin{equation}
\gamma\left(\mathbf{r}\tau,\mathbf{r}'\tau'\right)=\left\langle 
\exp\left\{ -i\left[\phi\left(\mathbf{r}\tau\right)
-\phi\left(\mathbf{r}'\tau'\right)\right]\right\} 
\right\rangle 
\end{equation}
 is the two-point phase correlator associated with the order parameter
field, where $\left\langle \dots\right\rangle $ is the averaging
with respect to the action:\begin{equation}
\mathcal{S}_{0}[\phi]=\sum_{{\bf r}}\int_{0}^{\beta}
d\tau\left[\frac{\dot{\phi}^{2}({\bf r}\tau)}{U}
+\frac{1}{i}\frac{2\mu}{U}\dot{\phi}({\bf r}\tau)\right].\end{equation}
The action with the topological contribution, after Fourier transform,
we write as 
\begin{equation}
\mathcal{S}_{\mathrm{eff}}[\bar{z},z]=\frac{1}{N\beta}\sum_{\mathbf{k},n}\bar{z}
\left(\mathbf{k}\omega_{n}\right)\mathrm{G}_{z0\mathbf{k}}^{-1}\left(\omega_{n}\right)
z\left(\mathbf{k}\omega_{n}\right),
\end{equation}
 where
\begin{equation}
\mathrm{G}_{z0\mathbf{k}}^{-1}\left(\omega_{n}\right)=\lambda_{z}
+\gamma^{-1}\left(\omega_{n}\right)\label{eq:ChargeSector_GreenFunction}
\end{equation}
is the inverse of the propagator, while the phase correlator
after Fourier transform, can be written as as: 
\begin{equation}
\gamma\left(\omega_{n}\right)=\frac{1}{\mathcal{Z}_{0}}\frac{4}{U}
\sum_{m=-\infty}^{+\infty}\frac{e^{-\frac{\beta U}{2}
\left(m+\frac{{\mu}}{U}\right)^{2}}}{1-4\left(m+\frac{{\mu}}{U}
-\frac{i\omega_{n}}{U}\right)^{2}},\label{correlator}
\end{equation}
 where 
\begin{equation}
\mathcal{Z}_{0}=\sum_{m=-\infty}^{+\infty}
\exp\left[-\frac{1}{2}\beta U\left(m+\frac{{\mu}}{U}\right)^{2}\right]
\label{statsuma}
\end{equation}
 is the partition function for the set of non-interacting quantum
rotors.  Note that the presence of the integer winding numbers
in Eqs (\ref{correlator}) and  (\ref{statsuma}) renders the phase propagator periodic
in the reduced chemical potential $\mu/U$.
The unimodular condition of the U(1) phase variables translates into the
equation
\begin{equation}
1=\frac{1}{N\beta}\sum_{\mathbf{k},n}\frac{1}{\lambda_{z0}+\gamma^{-1}
\left(\omega_{n}\right)},\label{critical line}
\end{equation}
which fixes the Lagrange multiplier $\lambda_{z0}$.	
%
\subsection{Fermionic action}
%
Now we turn to  the effective action of pseudo--fermions by tracing
out the gauge degrees of freedom. To this end we write the partition
function as 
\begin{eqnarray}
{\cal Z} & = & \int\left[{\cal D}\bar{h}{\cal D}{h}\right]e^{-{\cal S}[\bar{h},h]},
\end{eqnarray}
 where
\begin{eqnarray}
{\cal S}[\bar{h},h] & = & -\ln\int[{\cal D}\phi{\cal D}{\bf \Omega}]
e^{-{\cal S}[{\varphi,\phi,\vartheta},\bar{h},h]}\nonumber \\
 & = & \mathcal{S}_{B}[\bar{h},h]+{\cal S}_{t}^{(1)}[\bar{h},h]+{\cal S}_{t}^{(2)}[\bar{h},h]
\nonumber \\
 & + & \int_{0}^{\beta}d\tau{\cal H}_{{\bf \Omega,\phi}}^{(\rho)}
\left[\bar{h},h\right].
\end{eqnarray}
The kinetic part is calculated in the cumulant expansion. In the first
order of the expansion:
\begin{eqnarray}
\mathcal{S}_{t}^{(1)}\left[\bar{h},h\right] & = & \sum_{\langle{\bf r}{\bf r}'\rangle,\alpha}
\int_{0}^{\beta}d\tau\left(tg+\bar{\mu}\delta_{\mathbf{r}\mathbf{r}'}\right)\nonumber \\
 & \times & \left[\bar{h}_{{\alpha}}({\bf r}\tau)h_{\alpha}
({\bf r}'\tau)+h.c.\right].
\end{eqnarray}
The hopping $t$ is renormalized by a Gutzwiller--type parameter\cite{guz} $g  =  g_{c}g_{s}$,
where
\begin{eqnarray}
g_{s} & = & \langle\left[{\bf R}^{\dagger}({\bf r}\tau)
{\bf R}({\bf r'}\tau)\right]_{11}\rangle=
\langle[{\bf R}^{\dagger}({\bf r}\tau){\bf R}({\bf r'}\tau)]_{22}\rangle
\nonumber \\
g_{c} & = & \langle\bar{z}({\bf r}\tau)z({\bf r'}\tau)\rangle
\end{eqnarray}
being a multiply of the renormalization parameters in charge and spin sectors.
The parameters $g_s$ and $g_c$ have to be calculated  self-consistently,
they contribute to the band narrowing and eventually for the band collapse 
for strong correlations.
Another contribution to the hopping amplitude comes from the
the second order cumulant  expansion which generates a term of the form:
\begin{equation}
{\cal S}_{t}^{(2)}[\bar{h},h]=-\frac{2t^{2}}{U}\int_{0}^{\beta}d\tau
\sum_{\langle{\bf r}{\bf r}'\rangle}{\cal F}^{\dagger}({\bf r}\tau{\bf r}'\tau){\cal F}
({\bf r}\tau{\bf r}'\tau)
\end{equation}
 with the bond operators 
\begin{equation}
{\cal F}({\bf r}\tau{\bf r}'\tau)=\frac{\bar{h}_{\uparrow}({\bf r}\tau)
h_{\uparrow}({\bf r'}\tau)+\bar{h}_{\downarrow}({\bf r}\tau)h_{\downarrow}({\bf r'}\tau)}{\sqrt{2}}.
\end{equation}
The fourth-order fermionic operator terms in the action can be decoupled
using HS transform which introduces additional field $v$. The resulting
action becomes bilinear
\begin{equation}
{\cal S}_{t}^{(2)}[\bar{h},h]=t_{J}\sum_{\langle{\bf rr'}\rangle\alpha}
\int_{0}^{\beta}d\tau[\bar{h}_{\alpha}({\bf r}\tau)h_{\alpha}({\bf r'}\tau)+h.c.],
\end{equation}
with the effective hopping that is proportional to the antiferromagnetic exchange constant
\begin{equation}
t_{J}=\frac{Jv}{4}.
\end{equation}
This dispersive low-energy band, for which the band width is set by the exchange interaction
is a clear signature of the coupling of the quasiparticles to antiferromagnetic correlations.
The value of the $v$ field can be fixed self-consistently using saddle-point method
to give	
\begin{equation}
v=\sum_{\alpha}\langle\bar{h}_{\alpha}({\bf r}\tau)h_{\alpha}({\bf r'}\tau)\rangle.	
\end{equation}
As we will see in the following for a certain range of  model parameters the
value of $v$ may vanish leading to the band collapse and the insulating state.
Finally, one can write the resulting fermionic action in a compact
Nambu form:
\begin{equation}
{\cal S}[\bar{h},h]=\frac{1}{\beta N}\sum_{\mathbf{k},n}\bar{\Lambda}_{h}
\left(\mathbf{k}\omega_{n}\right)G_{h0\mathbf{k}}^{-1}\left(\omega_{n}\right)
\Lambda_{h}\left(\mathbf{k}\omega_{n}\right),\label{eq:FermionicSector_GreenFunction}
\end{equation}
where the vectors $ \Lambda_{h}\left(\mathbf{k}\tau\right)$ are defined by 
\begin{equation}
\Lambda_{h}\left(\mathbf{k}\tau\right)=\left[\begin{array}{c}
h_{\uparrow}\left(\mathbf{k}\omega_{n}\right)\\
h_{\downarrow}\left(\mathbf{k}\omega_{n}\right)\\
h_{\uparrow}\left(\mathbf{k}-\bm{\pi}\omega_{n}\right)\\
h_{\downarrow}\left(\mathbf{k}-\bm{\pi}\omega_{n}\right)\end{array}\right]
\end{equation}
and the inverse propagator reads
 \begin{equation}
G_{h0\mathbf{k}}^{-1}\left(\omega_{n}\right)=\left[\begin{array}{cccc}
\omega_{h\mathbf{k}}^{-} & 0 & \Delta_{c} & 0\\
0 & \omega_{h\mathbf{k}}^{-} & 0 & -\Delta_{c}\\
\Delta_{c} & 0 & \omega_{h\mathbf{k}}^{+} & 0\\
0 & -\Delta_{c} & 0 & \omega_{h\mathbf{k}}^{+}\end{array}\right]
\end{equation}
with $\omega_{h\mathbf{k}}^{\pm}=i\omega_{n}-\bar{\mu}\pm2\chi\xi_{\mathbf{k}}$,
where $\chi=t_J/2$.
%
\subsection{Spin-angular action}
%
Since we are interested in the magnetic properties of the system a
natural step is to obtain the effective action that involves the spin-directional
degrees of freedom ${\bf \Omega}$, which important fluctuations correspond
to rotations. This can be done by integrating out fermions: 
\begin{eqnarray}
 &  & {\cal Z}=\int\left[{\cal D}{\bf \Omega}\right]e^{-{\cal S}[{\bf \Omega}]},
\end{eqnarray}
 where 
\begin{eqnarray}
{\cal S}[{\bf \Omega}]=-\ln\int[{\cal D}\phi{\cal D}
\bar{h}{\cal D}h]e^{-{\cal S}[\varphi,\phi,\vartheta,\bar{h},h]}
\end{eqnarray}
 generates the cumulant expansions for the low energy action in the
form 
\begin{equation}
{\cal S}[{\bf \Omega}]=\mathcal{S}_{0}\left[\mathbf{\Omega}\right]+{\cal S}_{B}[{\bf \Omega}]
+{\cal S}_{J}[{\bf \Omega}].
\end{equation}
%
%
\subsubsection{AF exchange term}
%
The part of the action that involves the spin stiffnesses is given by
\begin{eqnarray}
{\cal S}_{J}[{\bf \Omega}] & = & \frac{J\left(\Delta\right)}{4}
\sum_{\langle{\bf r}{\bf r}'\rangle}\int_{0}^{\beta}d\tau
{\bf \Omega}({\bf r}\tau)\cdot{\bf \Omega}({\bf r'}\tau)
\label{eq:SJ}
\end{eqnarray}
 with the AF-exchange coefficient
 \begin{eqnarray}
 &  & J(\Delta_{c})=\frac{4t^{2}}{U}(n_{\uparrow}-n_{\downarrow})^{2}\equiv
\frac{4t^{2}}{U}\left(\frac{2\Delta_{c}}{U}\right)^{2}.\label{afex}
\end{eqnarray}
 From the Eq. (\ref{afex}) it is evident that for $U\to\infty$ one
has $J(\Delta_{c})\sim\frac{4t^{2}}{U}$ since $\frac{2\Delta_{c}}{U}\to1$
in this limit. In general the AF-exchange parameter persists as long
as the charge gap $\Delta_{c}$ exists. However, $J(\Delta_{c})$
diminishes rapidly in the $U/t\to0$ weak coupling limit.
%
\subsubsection{Berry term}
%
In general, in addition to the usual exchange term, the action describing
antiferromagnetic spin systems is expected to have a topological Berry
phase term 
\begin{eqnarray}
{\cal S}_{B}[{\bf \Omega}] & = & -2\sum_{{\bf rr'}}\int_{0}^{\beta}
d\tau{\bf A}({\bf r}\tau)\cdot\langle{\bf S}_{h}({\bf r'}\tau')\rangle,
\end{eqnarray}
 where
 \begin{eqnarray}
 &  & \langle S_{h}^{z}({\bf r}\tau)\rangle=\frac{\Delta_{c}}{U}\equiv \theta.
\end{eqnarray}
 In terms of angular variables, the Berry term becomes
\begin{equation}
{\cal S}_{B}[{\bf \Omega}]=\frac{\theta}{i}
\sum_{{\bf r}}\int_{0}^{\beta}d\tau\left[\dot{\varphi}({\bf r}\tau)
\cos\vartheta({\bf r}\tau)+\dot{\chi}({\bf r}\tau)\right].\label{sberry}
\end{equation}
 If we work in Dirac {}``north pole\char`\"{} gauge $\chi({\bf r}\tau)=-\varphi({\bf r}\tau)$
one recovers the familiar form 
\begin{equation}
{\cal S}_{B}[{\bf \Omega}]=\frac{\theta}{i}\sum_{{\bf r}}
\int_{0}^{\beta}d\tau\dot{\varphi}({\bf r}\tau)
[1-\cos\vartheta({\bf r}\tau)].
\end{equation}
Here, the integral of the first term in Eq. (\ref{sberry}) has a
simple geometrical interpretation as it is equal to a solid angle
swept by a unit vector ${\bf \Omega}(\vartheta,\varphi)$ during its
motion. The extra phase factor coming from the Berry phase, requires
some little extra care, since it will induce quantum mechanical phase
interference between configurations. In regard to the non-perturbative
effects, we realized the presence of an additional parameter with
the topological angle or so-called theta term 
 that is related to the Mott gap. In the large-$U$ limit one has
$\Delta_{c}\to U/2$, so that $\theta\to\frac{1}{2}$ relevant for
the half-integer spin. However, for arbitrary $U$ the theta term
will be different from that value, which, as we show, will be instrumental
for destruction of the antifferomagnetic order away from the spin-localized
$U\to\infty$ limit.
%
\subsubsection{Kinetic energy term for  spin}
In analogy to the charge U(1) field the SU(2) spin system exhibit
emergent dynamics. Integration of fermions will generate the kinetic
term for the SU(2) rotors
\begin{eqnarray}
\mathcal{S}_{0}\left[\mathbf{\Omega}\right] & = & 2\sum_{\mathbf{r}}
\int_{0}^{\beta}d\tau\left[\chi_{T}\sum_{a=x,y}A^{a}
\left(\mathbf{r}\tau\right)A^{a}\left(\mathbf{r}\tau\right)+\right.\nonumber \\
 & + & \left.\chi_{L}A^{z}\left(\mathbf{r}\tau\right)A^{z}\left(\mathbf{r}\tau\right)\right],
\end{eqnarray}
which can be written in a more compact form:
\begin{eqnarray}
\mathcal{S}_{0}\left[\mathbf{\Omega}\right] & = & 
2\sum_{\mathbf{r}}\int_{0}^{\beta}d\tau
\left[\chi_{T}\sum\mathbf{A}\left(\mathbf{r}\tau\right)\cdot\mathbf{A}
\left(\mathbf{r}\tau\right)+\right.
\nonumber \\
 & + & \left.\left(\chi_{L}-\chi_{T}\right)
A^{z}\left(\mathbf{r}\tau\right)A^{z}\left(\mathbf{r}\tau\right)\right],
\end{eqnarray}
where:
\begin{eqnarray}
\chi_{T} & = & \left\langle S_{h}^{x}\left(\mathbf{r}\tau\right)
S_{h}^{x}\left(\mathbf{r}\tau\right)\right\rangle =\left\langle 
S_{h}^{y}\left(\mathbf{r}\tau\right)S_{h}^{y}\left(\mathbf{r}\tau\right)\right\rangle ,
\nonumber \\
\chi_{L} & = & \left\langle S_{h}^{z}\left(\mathbf{r}\tau\right)
S_{h}^{z}\left(\mathbf{r}\tau\right)\right\rangle .
\end{eqnarray}
The transverse susceptibility behaves  in weak and strong coupling
limit as follows\cite{chubuk}
\begin{equation}
\chi_{T}\sim\begin{cases}
\frac{1}{8J} & \,\,\, t\ll U\\
\frac{1}{2\pi}\frac{1}{t}\sqrt{\frac{t}{U}} & \,\,\, t\gg U.
\end{cases}
\end{equation}
Thus, in the large--$U$ limit the spin kinetic part vanishes  and the
Hubbard model  maps, as expected,	 onto a quantum spin–-1/2
Heisenberg model with near– neighbor antiferromagnetic exchange integral $J = 4t^2/U$. 
Here, the superexchange interaction  $J$ is determined by the virtual hopping of an
electron of a given spin to an adjacent site containing an
electron with an opposite spin. Thus the dynamics of
$J$ involves virtual excitations above the Mott gap which
is set by $U$, and the effective interaction is essentially instantaneous.
However, the retarded contribution occurs on an energy scale which is
small compared to the bare bandwidth  and the onsite
Coulomb interaction. For the correlated systems, 
the relative weight of the retarded and nonretarded 
contributions to the effective  interaction remains an open question.

%
\subsubsection{CP$^{1}$ representation}
%
Now, we use a compact matrix notation adapted to the
SU(2)-invariant character of the Hamiltonian and the effective
action including a consistent scheme of coherent states
within a functional-integral formulation.	
It the CP$^{1}$ representation the spin-quantization axis, can be
conveniently written as
\begin{equation}
\mathbf{\Omega}\left(\mathbf{r}\tau\right)=\sum_{\alpha\alpha'}
\bar{\zeta}_{\alpha}\left(\mathbf{r}\tau\right)\bm{\sigma}_{\alpha\alpha'}
\zeta_{\alpha'}\left(\mathbf{r}\tau\right).
\end{equation}
Therefore, all the terms in the spin action can be expressed
as functions of  $\zeta_{\alpha}\left(\mathbf{r}\tau\right)$,
$\bar{\zeta}_{\alpha}\left(\mathbf{r}\tau\right)$
variables instead of angular variables, which are more complicated
to be handled. The spin--kinetic and Berry phase term now assume the simpler form
\begin{eqnarray}
 &  & {\bf A}({\bf r}\tau)\cdot{\bf A}({\bf r}\tau)=-\frac{1}{4}
\left[\dot{\vartheta}^{2}({\bf r}\tau)+\dot{\varphi}^{2}({\bf r}\tau)\right.\nonumber \\
 &  & \,\,\,\,\,+\left.\dot{\chi}^{2}({\bf r}\tau)+2\dot{\varphi}({\bf r}\tau)
\dot{\chi}({\bf r}\tau)\cos\vartheta({\bf r}\tau)\right]\nonumber \\
 &  & \,\,\,\,\,\equiv-\dot{\bm{\zeta}}\left(\mathbf{r}\tau\right)\cdot
\dot{\bm{\zeta}}\left(\mathbf{r}\tau\right),\nonumber \\
 &  & A_{z}\left(\mathbf{r}\tau\right)=\frac{i}{2}
\dot{\varphi}({\bf r}\tau)\cos\vartheta({\bf r}\tau)
+\frac{i}{2}\dot{\chi}({\bf r}\tau)\nonumber \\
 &  & \,\,\,\,\,\equiv\frac{1}{2}\left[\bar{\bm{\zeta}}
\left(\mathbf{r}\tau\right)\cdot\dot{\bm{\zeta}}
\left(\mathbf{r}\tau\right)-\dot{\bar{\bm{\zeta}}}
\left(\mathbf{r}\tau\right)\cdot\bm{\zeta}
\left(\mathbf{r}\tau\right)\right].
\label{phiaction}
\end{eqnarray}
Consequently, the spin--angular action transforms into 
\begin{eqnarray}
 &  & \mathcal{S}\left[\bar{\bm{\zeta}},\bm{\zeta}\right]
=\sum_{\mathbf{r}}\int_{0}^{\beta}d\tau\left\{ 2\chi_{T}
\dot{\bm{\zeta}}\left(\mathbf{r}\tau\right)\cdot\dot{\bm{\zeta}}
\left(\mathbf{r}\tau\right)\right.\nonumber \\
 &  & +\frac{\chi_{L}-\chi_{T}}{2}\left[\bar{\bm{\zeta}}
\left(\mathbf{r}\tau\right)\cdot\dot{\bm{\zeta}}
\left(\mathbf{r}\tau\right)-\dot{\bar{\bm{\zeta}}}
\left(\mathbf{r}\tau\right)\cdot\bm{\zeta}\left(\mathbf{r}\tau\right)\right]^{2}
\nonumber \\
 &  & -\left.\theta\left(-1\right)^{\mathbf{r}}\left[\bar{\bm{\zeta}}
\left(\mathbf{r}\tau\right)\cdot\dot{\bm{\zeta}}\left(\mathbf{r}\tau\right)
-\dot{\bar{\bm{\zeta}}}\left(\mathbf{r}\tau\right)\cdot\bm{\zeta}
\left(\mathbf{r}\tau\right)\right]\right\} 
\nonumber \\
 &  & -J\sum_{\left\langle \mathbf{r}\mathbf{r}'\right\rangle }
\int_{0}^{\beta}d\tau\bar{\mathcal{A}}\left(\mathbf{r}\tau\mathbf{r}'\tau\right)
\mathcal{A}\left(\mathbf{r}\tau\mathbf{r}'\tau\right)
\label{eq:Szetabzeta}
\end{eqnarray}
with the bond operators:
\begin{eqnarray}
 &  & \bar{\mathcal{A}}\left(\mathbf{r}\tau\mathbf{r}'\tau\right)
\mathcal{A}\left(\mathbf{r}\tau\mathbf{r}'\tau\right)
=-\frac{1}{4}\mathbf{\Omega}\left(\mathbf{r}\tau\right)
\cdot\mathbf{\Omega}\left(\mathbf{r}'\tau\right)+\frac{1}{4}
\nonumber \\
 &  & \mathcal{A}\left(\mathbf{r}\tau\mathbf{r}'\tau\right)
=\frac{\zeta_{\uparrow}\left(\mathbf{r}\tau\right)
\zeta_{\downarrow}\left(\mathbf{r}'\tau\right)-
\zeta_{\downarrow}\mathbf{\left(\mathbf{r}\tau\right)}
\zeta_{\uparrow}\left(\mathbf{r}'\tau\right)}{\sqrt{2}}
\end{eqnarray}
relevant for the bosonic representation of an antiferromagnet.\cite{auerbach}
%
\subsubsection{Canonical transformation of CP$^{1}$ variables}
%
In order to achieve a consistent representation of the underlying
antiferromagnetic structure, it is unavoidable to explicitly split
the degrees of freedom according to their location on sublattice A
or B. Since the lattice is bipartite allowing one to make the unitary
transformation\cite{auerbach}
\begin{eqnarray}
\zeta_{\uparrow}({\bf r}\tau) & \to & -\zeta_{\downarrow}({\bf r}\tau)
\nonumber \\
\zeta_{\downarrow}({\bf r}\tau) & \to & \zeta_{\uparrow}({\bf r}\tau)
\end{eqnarray}
 for sites on one sublattice, so that the antiferromagnetic bond operator
becomes
\begin{eqnarray}
{\cal A}({\bf r}\tau{\bf r'}\tau) & \to & {\cal A}'({\bf r}\tau{\bf r'}\tau)
=\sum_{\alpha=1}^{2}\frac{\zeta_{\alpha}({\bf r}\tau)\zeta_{\alpha}({\bf r}'\tau)}{\sqrt{2}}.
\end{eqnarray}
This canonical transformation preserves the constraint in Eq. (\ref{eq:CP1constraint}).
Biquadratic (four-variable) terms in the Lagrangian cannot be readily
integrated in the path integral. Introducing a complex variable for
each bond that depends on {}``imaginary time\char`\"{} $Q({\bf r}\tau{\bf r'}\tau)$
we decouple the four-variable terms 
$\bar{{\cal A}}'({\bf r}\tau{\bf r'}\tau){\cal A}'({\bf r}\tau{\bf r'}\tau)$.
 In a similar manner by introducing
a local real field $a\left(\mathbf{r}\tau\right)$, we can decouple
second term in the r.h.s. in the Eq. (\ref{eq:Szetabzeta}). To handle
the unimodularity condition one introduces a Lagrange multiplier $\lambda_{\zeta}(\tau)$.
to treat the variables $\zeta_{\alpha}({\bf r}\tau)$, $\bar{\zeta}_{\alpha}({\bf r}\tau)$
as unconstrained bosonic fields. Consequently the effective Hamiltonian becomes
\begin{eqnarray}
{\cal H}_{Q}[\bar{\bm{\zeta}},{\bm{\zeta}}] & = & 
\sum_{\langle{\bf r}{\bf r}'\rangle}\int_{0}^{\beta}
d\tau\left[a\left(\bar{\bm{\zeta}}\cdot\dot{\bm{\zeta}}
-\dot{\bar{\bm{\zeta}}}\cdot\bm{\zeta}\right)\delta_{\mathbf{r}\mathbf{r}'}\right.
\nonumber \\
 & + & \left.Q{\bar{\bm{\zeta}}}\cdot\bar{\bm{\zeta}}
+\bar{Q}{\bm{\zeta}}\cdot{\bm{\zeta}}+\lambda_{\zeta}{\bar{\bm{\zeta}}}
\cdot{\bm{\zeta}}\delta_{\mathbf{r}\mathbf{r}'}\right]
\end{eqnarray}
where
\begin{equation}
\tilde{\chi}=\frac{\chi_{T}-\chi_{L}}{2}.
\end{equation}
The saddle-point values of the $Q$, $a$
and $\lambda_{\zeta}$ fields are given by
\begin{eqnarray}
a_{sp}\left(\mathbf{r}\tau\right) & = & \tilde{\chi}\left\langle 
\bar{\bm{\zeta}}\cdot\dot{\bm{\zeta}}-\dot{\bar{\bm{\zeta}}}
\cdot\bm{\zeta}\right\rangle =0\nonumber \\
Q_{{\rm sp}}({\bf r}\tau{\bf r}'\tau) & = & -\frac{J}{2}\langle
\bar{\bm{\zeta}}({\bf r}\tau)\cdot\bar{\bm{\zeta}}({\bf r}'\tau)\rangle
\nonumber \\
1 & = & \langle\bar{\bm{\zeta}}({\bf r}\tau)\cdot\bm{\zeta}({\bf r}\tau)
\rangle
\end{eqnarray}
and by assuming the uniform solutions $Q_{{\rm sp}}({\bf r}\tau{\bf r}'\tau)\equiv Q$,
$a_{sp}\left(\mathbf{r}\tau\mathbf{r}'\tau\right)\equiv a$ 
and $\lambda_{\zeta sp}\left(\tau\right)\equiv\lambda_{\zeta}$
we obtain for the Hamiltonian in the spin-bosonic sector
\begin{equation}
\mathcal{H}\left[\bar{\bm{\zeta}},\bm{\zeta}\right]
=\frac{1}{2\beta N}\sum_{\mathbf{k}n\sigma}
\bar{\Lambda}_{\zeta\sigma}\left(\mathbf{k}\omega_{n}\right)
G_{\zeta0\mathbf{k}}^{-1}\left(\omega_{n}\right)
\Lambda_{\zeta\sigma}\left(\mathbf{k}\omega_{n}
\right)\label{eq:SpinSector_finalaction}
\end{equation}
with
\begin{equation}
\Lambda_{\zeta\sigma}\left(\mathbf{k}\omega_{n}\right)=\left[\begin{array}{c}
\zeta_{\sigma}\left(\mathbf{k},\omega_{n}\right)\\
\bar{\zeta}_{\sigma}\left(-\mathbf{k},-\omega_{n}\right)\\
\zeta_{\sigma}\left(\mathbf{k}-\bm{\pi},\omega_{n}\right)\\
\bar{\zeta}_{\sigma}\left(-\mathbf{k}+\bm{\pi},-\omega_{n}\right)\end{array}\right]
\end{equation}
and
\begin{equation}
G_{\zeta0\mathbf{k}}^{-1}\left(\omega_{n}\right)=\left[\begin{array}{cccc}
\frac{\omega_{n}^{2}}{\mathcal{E}_{s}}+\lambda_{\zeta}
 & 2Q\xi_{\mathbf{k}} & -2i\theta\omega_{n} & 0\\
2Q\xi_{\mathbf{k}} & \frac{\omega_{n}^{2}}{\mathcal{E}_{s}}
+\lambda_{\zeta} & 0 & 2i\theta\omega_{n}\\
-2i\theta\omega_{n} & 0 & \frac{\omega_{n}^{2}}{\mathcal{E}_{s}}
+\lambda_{\zeta} & -2Q\xi_{\mathbf{k}}\\
0 & 2i\theta\omega_{n} & -2Q\xi_{\mathbf{k}} &
	 \frac{\omega_{n}^{2}}{\mathcal{E}_{s}}+\lambda_{\zeta}
\end{array}\right]
\end{equation}
and $\mathcal{E}_{s}=1/\left(2\chi_{T}\right)$ sets the kinetic energy scale for the
SU(2) rotors. 
%
\section{Self-Consistent Equations\label{sec:Self-Consistent-Equations}}
%
Procedure of decoupling of the Hubbard Hamiltonian in Eq. (\ref{mainham})
introduces numerous decoupling fields, which values are fixed within
saddle-point approximation. Also, the bosonic degrees of freedom in
charge and spin sectors can condense leading to superconducting and
magnetic ordering (although in the present paper we only consider
magnetic ordering within the spin sector). Since, the ordering is
described within quantum rotor  model,
it introduces additional constraints in each sectors for order parameters
or Lagrange multipliers. Together, it creates a set of non-linear
self-consistent equations, which allow for calculation of the effective
variables of the present theory. 
%
\subsection{Charge sector}
%
With the charge sector Green's function from Eq. (\ref{eq:ChargeSector_GreenFunction})
one can write the constraint for  Lagrange multiplier
$\lambda_{z}$:
\begin{equation}
1=\frac{1}{\beta}\sum_{n}G_{z}\left(\omega_{n}\right)
=\frac{\coth\left(\frac{\beta E_{c}^{+}}{2}\right)
-\coth\left(\frac{\beta E_{c}^{-}}{2}\right)}
{4\sqrt{f\left(\frac{2\mu}{U}\right)^{2}
+\frac{2\delta\lambda_{z}}{U}}},\label{eq:Constraint_z}
\end{equation}
where
\begin{equation}
E_{c}^{\pm}=\frac{U}{2}\left[-f\left(\frac{2\mu}{U}\right)\pm\sqrt{f\left(\frac{2\mu}{U}\right)^{2}
+\frac{2\delta\lambda_{z}}{U}}\right]
\end{equation}
with $f\left(2\mu/U\right)=\mathrm{frac}\left(2\mu/U-1/2\right)-1/2,$ where
$\mathrm{frac}\left(x\right)=x-\left[x\right]$ is a fractional part
of $x$, since for temperatures much lower than $U$ ($\beta\ll U$)
the summation over winding numbers leads to periodic dependence of
the model on the chemical potential. 
%
%
\subsection{Fermionic sector}
The fermionic sector introduces the Mott-charge gap $\Delta_{c}$ and
the $v$ field, which renormalizes hopping in the second order of
cumulant expansion:
\begin{eqnarray}
\Delta_{c} & = & \frac{U}{\beta N}\sum_{\mathbf{k},n}
\left[G_{h\mathbf{k}}^{11}\left(\nu_{n}\right)-G_{h
\mathbf{k}-\bm{\pi}}^{22}\left(\nu_{n}\right)\right]
\nonumber \\
 & = & \frac{U}{N}\sum_{\mathbf{k}}\frac{\Delta_{c}}
{2E_{\mathbf{k}}}\left[n_{F}\left(E_{\mathbf{k}}^{-}\right)
-n_{F}\left(E_{\mathbf{k}}^{+}\right)\right],
\nonumber \\
v & = & \frac{1}{\beta N}\sum_{\mathbf{k},n}
\sum_{\alpha}F_{h}^{\alpha\alpha}\left(\mathbf{k}\nu_{n}\right)
\nonumber \\
 & = & \frac{J}{z}\frac{v}{N}\sum_{\mathbf{k}}
\frac{\xi_{\mathbf{k}}^{2}}{E_{\mathbf{k}}}\left[n_{F}
\left(E_{\mathbf{k}}^{-}\right)-n_{F}\left(E_{\mathbf{k}}^{+}\right)\right],
\end{eqnarray}
where $n_{F}\left(E\right)$ is the Fermi distribution,
\begin{equation}
E_{\mathbf{k}}^{\pm}=-\bar{\mu}\pm E_{\mathbf{k}}\mathrm{, with~}E_{\mathbf{k}}=\sqrt{\Delta_{c}^{2}
+\left(\frac{J}{2}v\xi_{\mathbf{k}}\right)^{2}}
\end{equation}
and the lattice structure factor 
\begin{equation}
\xi_{\mathbf{k}}=\cos k_{x}+\cos k_{y}.
\end{equation}
The presence of  $\Delta_c$ gives rise to a Fermi surface instability 
as first suggested by Slater.\cite{slater}
The basic principle behind an antiferromagnetic Slater insulator can be
explained most easily by electrons living on  a bipartite lattice.
that could be separated into two inter-penetrating sublattices (let say, A and B)
such that the nearest neighbor of any site are members of the opposite sublattice.
In the corresponding band structure picture the lattice unit cell is doubled
and the first Brillouin zone is cut in half.
%
\subsection{Spin sector}
%
In the Hubbard model the quantum-mechanical objects are not local spins
but mobile electrons such that we have to expect that the
analysis of the ground-state phase diagram as a function of the 
interaction strength is even more difficult than for the Heisenberg model.
So far in the above discussion  we did not consider
the possibility of an ordering of magnetic moments. Thus, the concept of
the Slater insulator has to be supplemented by that of the Mott–-Heisenberg insulator 
  displays long-range order. Since correlations are absent in a Hartree--Fock 
description the moments order at
the very same temperature they are formed. In contrast, the moments are already
 present in the Mott–--Heisenberg insulating state and remain in the paramagnetic  phase.
Therefore the ordering  of the  pre-formed moments, the important signature of
electron correlations,
provide a clear distinction between the ideas of Slater (self-consistent 
single-electron theory) and Mott (many-electron correlations).
Therefore, a nonzero value of $\Delta_{c}$ does not
imply the existence of AF long--range order. For this the angular
degrees of freedom ${\bf \Omega}({\bf r}\tau)$ have also to be ordered,
whose low-lying excitations are in the form of spin waves. In the
CP$^{1}$ representation (where the Neel field is represented by two
Schwinger bosons) Bose-Einstein condensation of the Schwinger bosons
at zero temperature signals the appearance of AF long-range order.
The AF order parameter in terms of the original fermion operators
is defined as 
\begin{eqnarray}
m_{AF} & = & \sum_{{\bf r}}(-1)^{{\bf r}}\langle S^{z}({\bf r}\tau)\rangle=
\nonumber \\
 & = & \sum_{{\bf r}}(-1)^{{\bf r}}\langle{\bf \Omega}
({\bf r}\tau)\rangle\cdot\langle{\bf S}_{h}({\bf r}\tau)\rangle.
\end{eqnarray}
 Owing the fact that $\langle S_{h}^{a}({\bf r}\tau)\rangle=(-1)^{{\bf r}}\Delta_{c}\delta_{a,z}$
we obtain 
\begin{eqnarray}
m_{AF} & = & \Delta_{c}\sum_{{\bf r}}\langle{\Omega}^{z}({\bf r}\tau)\rangle
\nonumber \\
 & = & \Delta_{c}\sum_{{\bf r}}
\left[\langle{\bar{\zeta}}_{\uparrow}
({\bf r}\tau){\zeta}_{\uparrow}({\bf r}\tau)\rangle
-\langle{\bar{\zeta}}_{\downarrow}({\bf r}\tau){\zeta}_{\downarrow}({\bf r}\tau)
\rangle\right].
\end{eqnarray}
 Furthermore, the order parameter for the CP$^{1}$ {}``boson condensate\char`\"{}
is\cite{auerbach}
\begin{eqnarray}
\langle{\bar{\zeta}}_{\alpha}({\bf k}\omega_{n})\rangle 
& = & \langle{\zeta}_{\alpha}({\bf k}\omega_{n})\rangle
\nonumber \\
 & = & \sqrt{\frac{\beta N}{2}}m_{0}\delta_{0,\omega_{n}}
\delta_{\uparrow,\alpha}\left(\delta_{{\bf k},0}+\delta_{{\bf k},{\bm \pi}}\right).
\end{eqnarray}
 This yields a macroscopic contribution (i.e., order one) to the staggered
magnetization and represents a macroscopic contribution to the CP$^{1}$
bosons density $m_0$, of the $\alpha=\uparrow$ bosons at the mode with
${\bf k}=0,\omega_{n}=0$ thus giving
 \begin{eqnarray}
m_{AF} & = & \frac{\Delta_{c}}{\beta UN}\sum_{{\bf k},\omega_{n}}
\left[\langle{\bar{\zeta}}_{\uparrow}({\bf k}\omega_{n}){\zeta}_{\uparrow}({\bf k}\omega_{n})
\rangle\right.\nonumber \\
 &  & \left.-\langle{\bar{\zeta}}_{\downarrow}({\bf k}
\omega_{n}){\zeta}_{\downarrow}({\bf k}\omega_{n})\rangle\right]=\frac{\Delta_{c}}{U}m_{0}^{2},
\end{eqnarray}
where the equation fixing  the order
parameter ${m}_0^{2}$ reads
\begin{eqnarray}
 &  & 1-{m}_0^{2}=\frac{1}{\beta N}\sum_{\mathbf{k}n\sigma}
G_{\zeta\mathbf{k}}^{\sigma}\left(\omega_{n}\right)
\nonumber \\
 &  & =\frac{1}{N}\sum_{\mathbf{k}}
\left\{ \frac{\coth\left[\beta E_{s\mathbf{k}}^{-}
\left(\omega_{\mathbf{k}}^{-}\right)\right]+\coth\left[\beta E_{s\mathbf{k}}^{+}
\left(\omega_{\mathbf{k}}^{-}\right)\right]}{4\sqrt{\theta^{2}+
\frac{\omega_{\mathbf{k}}^{-}}{\mathcal{E}_{s}}}}\right.
\nonumber \\
 &  & \left.+\frac{\coth\left[\beta E_{s\mathbf{k}}^{-}\left(\omega_{\mathbf{k}}^{+}\right)\right]
+\coth\left[\beta E_{s\mathbf{k}}^{+}\left(\omega_{\mathbf{k}}^{+}\right)\right]}{4\sqrt{\theta^{2}
+\frac{\omega_{\mathbf{k}}^{+}}{\mathcal{E}_{s}}}}\right\} .
\label{eq:Constraint_zeta}
\end{eqnarray}
Also, decoupling of the bond operators in the kinetic term of the
spin action in Eq. (\ref{eq:Szetabzeta}) leads to additional field
$Q$, which value is determined from the equation:
\begin{eqnarray}
 &  & Q=\frac{2J}{z\beta N}\sum_{\mathbf{k},n}\xi_{\mathbf{k}}
F_{\xi\mathbf{k}}\left(\omega_{n}\right)
\nonumber \\
 &  & =\frac{J}{zN}\sum_{\mathbf{k}}\xi_{\mathbf{k}}
\left\{ \frac{\coth\left[\beta E_{s\mathbf{k}}^{-}
\left(\omega_{\mathbf{k}}^{-}\right)\right]+
\coth\left[\beta E_{s\mathbf{k}}^{+}\left(\omega_{\mathbf{k}}^{-}\right)\right]}
{4\sqrt{\theta^{2}+\frac{\omega_{\mathbf{k}}^{-}}{\mathcal{E}_{s}}}}\right.
\nonumber \\
 &  & \left.+\frac{\coth\left[\beta E_{s\mathbf{k}}^{-}
\left(\omega_{\mathbf{k}}^{+}\right)\right]+\coth\left[\beta E_{s\mathbf{k}}^{+}
\left(\omega_{\mathbf{k}}^{+}\right)\right]}{4\sqrt{\theta^{2}+
\frac{\omega_{\mathbf{k}}^{+}}{\mathcal{E}_{s}}}}\right\} ,
\end{eqnarray}
where
\begin{equation}
E_{s\mathbf{k}}^{\pm}\left(\omega_{\mathbf{k}}^{\pm}\right)
=\frac{\mathcal{E}_{s}}{2}\left(\sqrt{\theta^{2}+
\frac{\omega_{\mathbf{k}}^{\pm}}{\mathcal{E}_{s}}}\pm\theta\right)
\end{equation}
and
\begin{equation}
\omega_{\mathbf{k}}^{\pm}=\lambda_{\zeta}\pm2Q\xi_{\mathbf{k}}.
\end{equation}
From the Eq.(\ref{eq:Constraint_zeta}) it follows that the magnetic 
order at finite temperatures is excluded in  two dimensions 
in agreement with Mermin–-Wagner theorem.\cite{mw}
%
\begin{figure}
\includegraphics[scale=0.7]{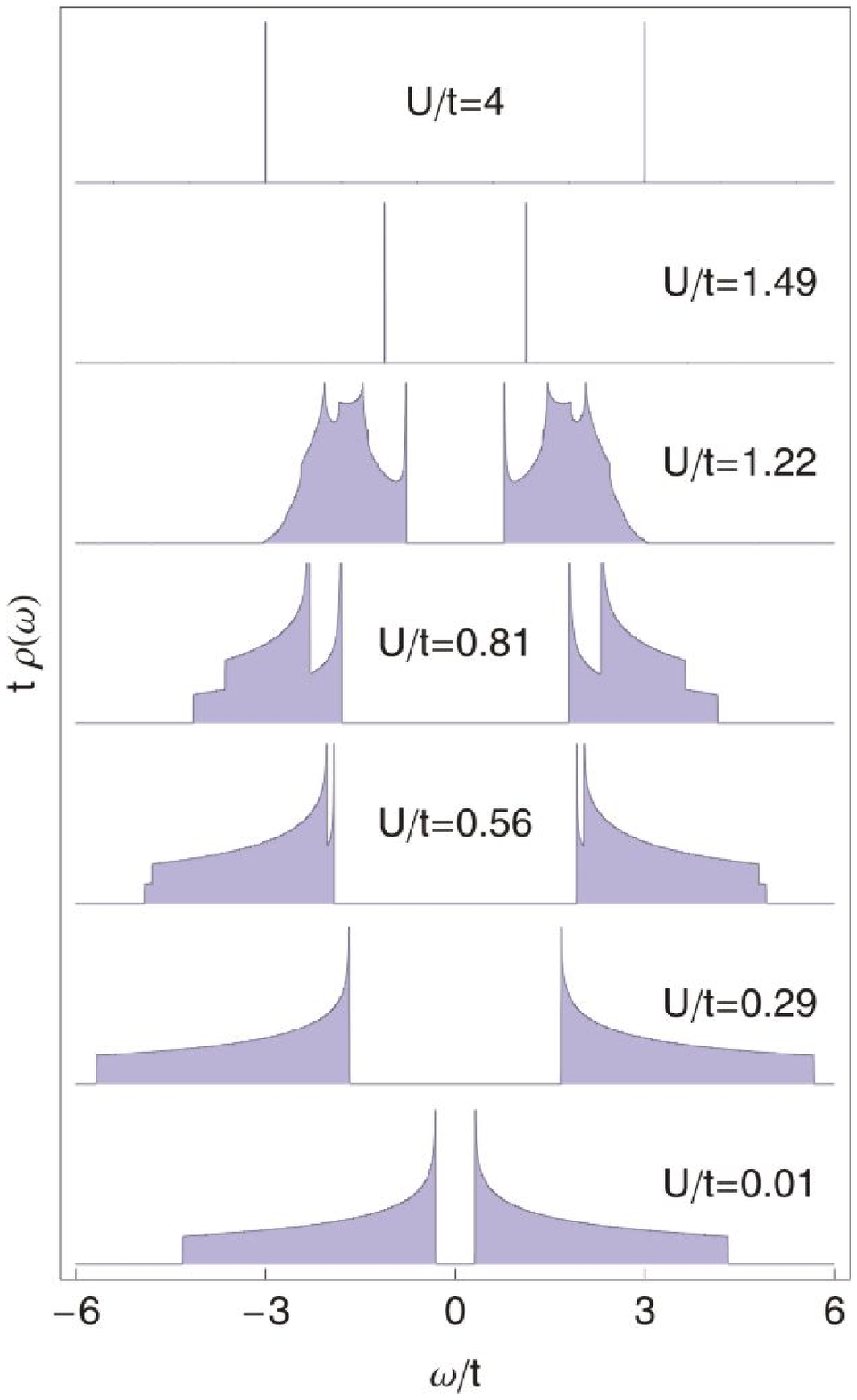}
\caption{(Color online) Evolution of the density of states of the model for various interactions
$U/t$ from strong to weak-coupling  limit. }
\end{figure}
%

\begin{figure}
\includegraphics[scale=0.7]{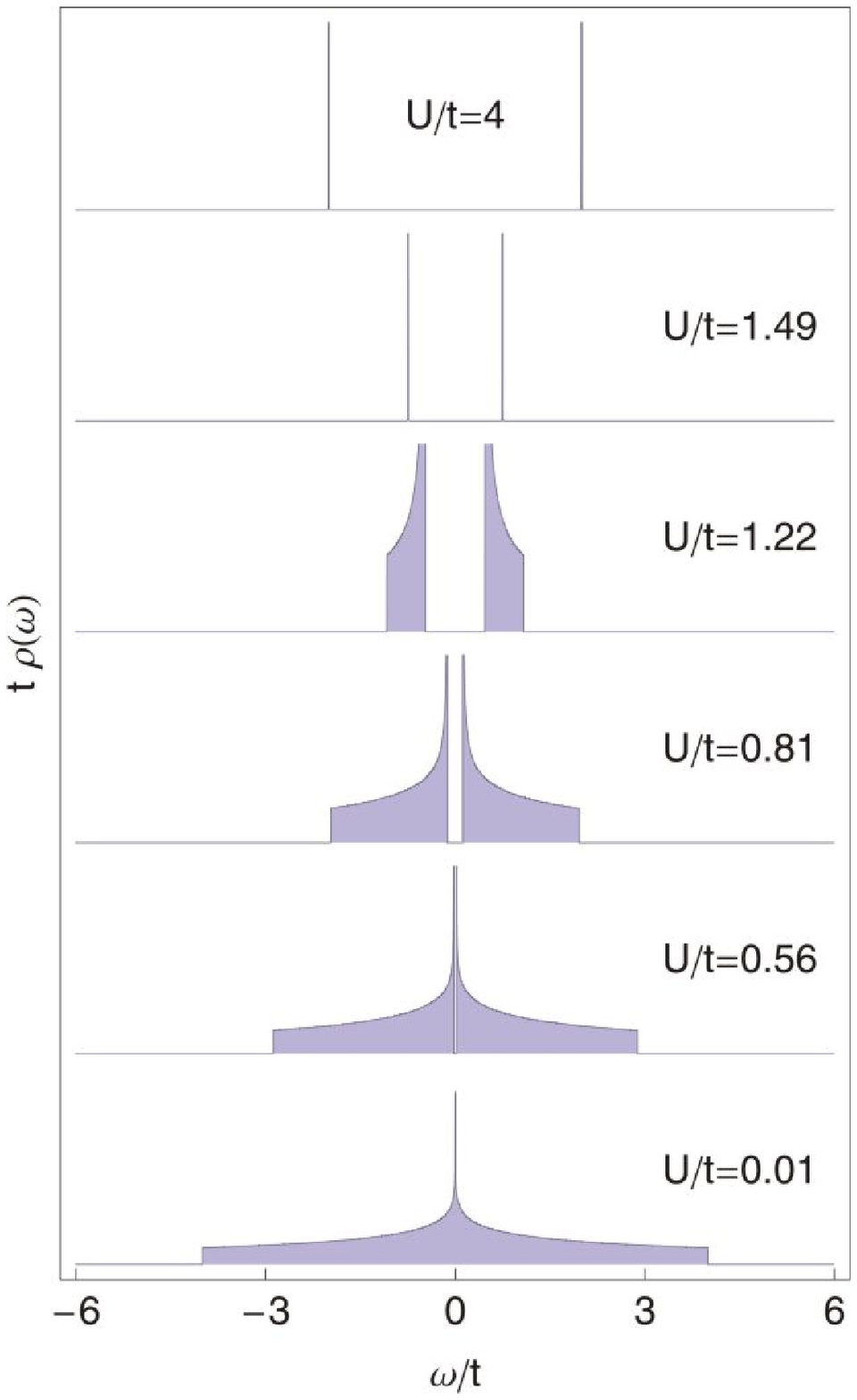}
\caption{(Color online) Evolution of the pseudo-fermionic density of states [see, Eq. (\ref{fermionic_density_states})]
of the model for various interactions $U/t$ from strong to weak-coupling  limit.}
\end{figure}
%

%
\begin{figure}
\includegraphics[scale=0.7]{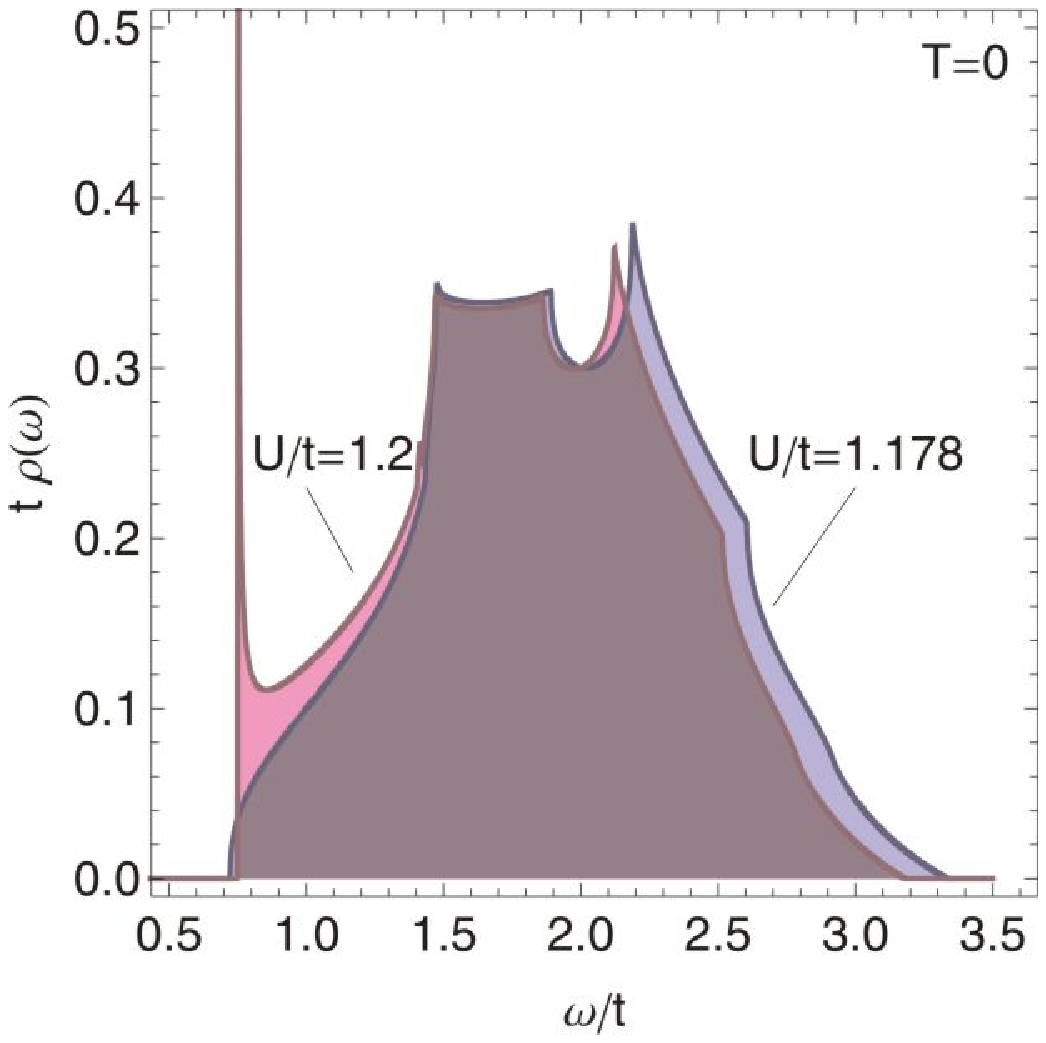}
\caption{(Color online) Spectral weight relocation induced by change of $U/t$ interaction
at zero temperature.
The weight of coherence peak (equal to\textbf{ ${m}_0^{2}$})
present for $U/t=1.2$ spectrum is relocated to the higher frequencies
for $U/t=1.178$, so the norm (spectrum integrand over frequency)
is preserved and equal to $1$.}
\end{figure}
%

%
\section{Single-Particle Spectral Functions}
%
We turn our attention now to the spectral function
defined in terms of the electron Green's function.
Within our construction, it is possible to write the electron
Green's function as a product of U(1) phase, SU(2) spin (in CP$^{1}$ representation) and  pseudo-fermion
Green's functions:
\begin{eqnarray}
G_{z}\left(\mathbf{r}\tau,\mathbf{r}'\tau'\right) 
& = & 
-\left\langle z\left(\mathbf{r},\tau\right)\bar{z}\left(\mathbf{r}',\tau'\right)\right\rangle ,
\nonumber \\
G_{\zeta}^{\alpha\alpha'}\left(\mathbf{r}\tau,\mathbf{r}'\tau'\right)
 & = & 
-\left\langle \zeta_{\alpha}\left(\mathbf{r},\tau\right)\bar{\zeta}_{\alpha'}
\left(\mathbf{r}',\tau'\right)\right\rangle ,\nonumber \\
G_{h}^{\alpha\alpha'}\left(\mathbf{r}\tau,\mathbf{r}'\tau'\right) 
& = & -\left\langle h_{\alpha}\left(\mathbf{r},\tau\right)\bar{h}_{\alpha'}
\left(\mathbf{r}',\tau'\right)\right\rangle ,
\label{vgre}
\end{eqnarray}
while the full Green's function of the system is the product:
\begin{eqnarray}
G_{\alpha\alpha}\left(\mathbf{r}\mathbf{r}'\tau\right) 
& = & -\sum_{\beta\gamma}\left\langle z_{\mathbf{r}}
\bar{z}_{\mathbf{r}'}\right\rangle \left\langle R_{\alpha\beta}
\left(\mathbf{r}\tau\right)R_{\gamma\alpha}^{\dagger}\left(\mathbf{r}'\tau\right)\right\rangle 
\nonumber \\
 & \times & \left\langle h_{\beta}\left(\mathbf{r}\tau\right)
h_{\gamma}\left(\mathbf{r}'\tau\right)\right\rangle .
\label{eq:FullGreenFunction}
\end{eqnarray}
%
%
\begin{figure}
\includegraphics[scale=0.7]{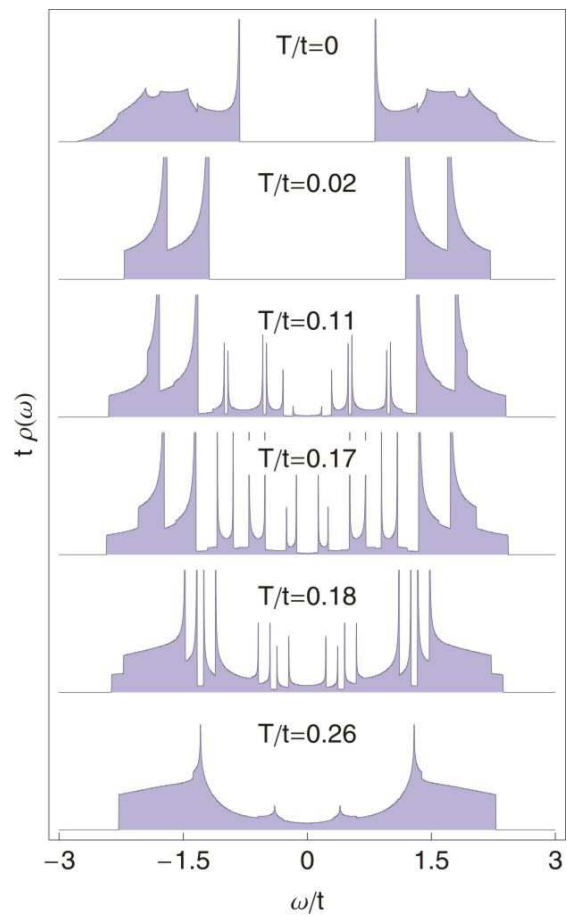}
\caption{(Color online) Evolution of density of states as a function of the temperature
(as indicated on the plot) 
for of $U/t=1.25$. }
\end{figure}
%
The problem of calculating
the spectral line shapes
now becomes one of calculating the convolution of the 
Green's functions in Eq.(\ref{vgre}).
Since, in the  antiferromagnetic phase 
 we allow for ordering in the spin sector, the averages over
spin variables can be non-zero  since CP$^1$ Bose condensation signals the appearance
of AF long-range order
\begin{eqnarray}
\left\langle \bar{\zeta}_{\alpha}\left(\mathbf{k},
\omega_{n}\right)\right\rangle &=&\left\langle 
\zeta_{\alpha}\left(\mathbf{k},\omega_{n}\right)\right\rangle
 \nonumber \\
&=&\sqrt{\frac{\beta N}{2}}{m_0}\delta_{0,\omega_{n}}
\delta_{\uparrow,\alpha}\left(\delta_{\mathbf{k},0}+
\delta_{\mathbf{k},\bm{\pi}}\right),
\end{eqnarray}
where the order parameter ${m_0}$ measures the fraction of the condensed
CP$^1$ bosons.
Consequently, the spin Green's function can be split into two contributions
according to
\begin{eqnarray}
 &  & G_{\zeta\mathbf{k}}^{\alpha\alpha}\left(\omega_{n}\right)
=-\left\langle \zeta_{\alpha}\left(\mathbf{k}\omega_{n}\right)
\bar{\zeta}_{\alpha}\left(\mathbf{k}\omega_{n}\right)\right\rangle 
\nonumber \\
 &  & =-\left[\left\langle \zeta_{\alpha}\left(\mathbf{0}0\right)
\bar{\zeta}_{\alpha}\left(\mathbf{0}0\right)\right\rangle 
+\left\langle \zeta_{\alpha}\left(\bm{\pi}0\right)\bar{\zeta}_{\alpha}
\left(\bm{\pi}0\right)\right\rangle \right]\delta_{\omega_{n}0}
\delta_{\alpha\uparrow}
\nonumber \\
 &  & -\left[1-\left(\delta_{\mathbf{k}0}+\delta_{\mathbf{k}\bm{\pi}}\right)
\delta_{\omega_{n}0}\delta_{\alpha\uparrow}\right]\left\langle \zeta_{\alpha}
\left(\mathbf{k}\omega_{n}\right)
\bar{\zeta}_{\alpha}\left(\mathbf{k}\omega_{n}\right)\right\rangle .
\end{eqnarray}
Because, in this case the average is periodic with respect to $\mathbf{k}$:
\begin{equation}
\left\langle \zeta_{\alpha}\left(\mathbf{k}\omega_{n}\right)
\bar{\zeta}_{\alpha}\left(\mathbf{k}\omega_{n}\right)\right\rangle
 =\left\langle \zeta_{\alpha}\left(\mathbf{k+\bm{\pi}}
\omega_{n}\right)\bar{\zeta}_{\alpha}\left(\mathbf{k+\bm{\pi}}\omega_{n}\right)
\right\rangle ,
\end{equation}
the spin sector Green's function reads:
\begin{eqnarray}
G_{\zeta\mathbf{k}}^{\alpha\alpha}\left(\omega_{n}\right)
 & = & -\beta N{m_0}^{2}\delta_{\alpha\uparrow}\delta_{\mathbf{k}0}\delta_{\omega_{n}0}
\nonumber \\
 & + & 
\left(1-2\delta_{\alpha\uparrow}\delta_{\mathbf{k}0}\delta_{\omega_{n}0}\right)
G_{\zeta\mathbf{k}}^{\alpha\alpha}\left(\omega_{n}\right).
\label{eq:SpinSector_GreenFunction}
\end{eqnarray}
Substituting this result into Eq. (\ref{eq:FullGreenFunction}) one can calculate
spectral density of the system (for details, see Appendix \ref{sec:Appendix_Spectral-densities}):
\begin{eqnarray}
 &  & A_{c\mathbf{k}}^{\alpha\alpha}\left(\omega\right)
=m_0^{2}A_{zh\mathbf{k}}^{\alpha\alpha}\left(\omega\right)\nonumber \\
 &  & \,\,\,\,+\frac{2}{N}\sum_{\mathbf{q}\ne0}
\int_{-\infty}^{+\infty}\frac{d\omega'}{2\pi}
A_{zh\mathbf{k}-\mathbf{q}}^{\alpha\alpha}\left(\omega-\omega'\right)
A_{\zeta\mathbf{q}}^{\alpha\alpha}\left(\omega'\right)\nonumber \\
 &  & \,\,\,\,\times\left[n_{B}\left(-\omega+\omega'\right)
+n_{F}\left(\omega'\right)\right]\nonumber \\
 &  & \,\,\,\,=m_0^{2}A_{zh\mathbf{k}}^{\alpha\alpha}
\left(\omega\right)+\sum_{\sigma}A_{zh\zeta\mathbf{k}}^{\sigma\sigma}\left(\omega\right)
\label{eq:SpectralDensity_Full}
\end{eqnarray}
and the density of states:
\begin{equation}
\rho_{c}^{\alpha\alpha}\left(\omega\right)
=m_0^{2}\rho_{zh}^{\alpha\alpha}\left(\omega\right)
+2\rho_{zh\zeta}^{\alpha\alpha}\left(\omega\right).
\end{equation}
The single electron density of states
contains then two terms. The first generates the 
the coherence peak associated with the log-range AF order, 
 which is a product of the condensate density and the pseudo-fermion spectral function.
To the extent that the peak and the background are distinguishable
objects, see Fig. 2, the weight under this quasi-particle
peak should be proportional to the condensate density of the CP$^1$ bosons.
Additionally, we have calculated the density of states corresponding only to fermionic
propagator $G_{h}^{\alpha\alpha}(\mathbf{k}\nu_{n})$ {[}see, Eq.
(\ref{fermionic_density_states}){]} to illustrate the correspondence
between Hartree-Fock approach and our method. The outcome is presented
in Fig. 3: the charge gap is a monotonic function of interaction strength
$U$, which is the feature of Hartree-Fock approaches.\cite{schrieffer}
Upon crossing the antiferromagnetic 
phase boundary, one observes  a remarkable  feature in the spectral density plot.
Figure 4 shows  how the spectrum  evolves
upon entering the ordered magnetic state as a function of the Coulomb interaction.
In a conventional Fermi  liquid system,
the spectrum would contain a narrow peak,
which is the signature of a well-defined
quasiparticle. In the AF state  exactly such a
peak is seen. As Fig. 4 shows, the
quasiparticle peak disappears upon leaving of the ordered state.
At the same time  the gap is preserved, yet the narrow peak
is gone.  As the peak disappears the spectral weight is transferred 
to the incoherent excitation background.
Evidently, the quasiparticle owes
its existence to the  of the condensation of the CP$^1$ bosons in the 
antiferromagnetic state, and not the energy
gap.
This exhibits a great similarity to that, which is seen in the ARPES
spectra in the underdoped samples, however in this case the condensate density
refers to the superconductor.
The electron spectral function is very broad above the superconducting
transition, but a sharp quasiparticle peak develops at the lowest
binding energies, followed by a dip and a broader hump,
giving rise to the so-called peak-dip-hump structure,\cite{damascelli} which is very 
similar to our results depicted in Fig. 4, where
the spectral density and density of states are sums of a coherent
part consisting of convolved functions, proportional
to the order parameter and incoherent -- being a convolution
of charge, pseudo-fermion and spin functions. 
It is interesting to note that the evolution
of a Mott-Hubbard insulator into a correlated metal  has been examined in the two-dimensional Hubbard
model by using the cellular dynamical mean-field theory,\cite{kyung} 
which incorporates short-range spatial correlations.
At half filling these correlations create  additional bands due to the ordered 
antiferromagnetic states that bear similarity to our results.
As far as the comparison with earlier works based on perturbation theory\cite{zlatic,lamas} is concerned, 
the main effect of correlations in the form of a transfer of the spectral weight to the high energies is reproduced.
The evolution of the spectral density
as a function of temperature  is depicted in Fig. 5. At finite temperature there is no
AF ordering according to the Mermin-Wagner theorem and consequently no coherence peak,
but one observes the gap filling as the temperature increases.
%
\section{Conclusions}
%
In this paper we have presented a method of calculation of spectral densities
for strongly correlated systems in terms of a collective
phase variable, the rotating quantization axis and the fermionic degrees of freedom.
In systems with strong Coulomb interactions,
the phase variable dual to the local charge is an important
collective field. 
A theory of the Hubbard model involving free fermionic degrees of freedom self consistently
coupled to a quantum U(1) and SU(2) rotor model has been developed.
The most interesting aspect of our approach lies however
in the possibility of going beyond purely local mean-field
description by incorporating the effect of spatial correlations and in particular
the influence of the ordered states on the spectral properties of the system.
This is very similar to the method implemented in Refs. \onlinecite{borejsza1,dupuis1},
where the single-particle properties are obtained by
writing the fermion field in terms of a Schwinger boson and
a pseudofermion whose spin is quantized along the fluctuating
N\'{e}el field. However, in the above works the charge sector was treated
on the mean-field level only and the lattice was approximated by the continuous long-wave limit. 
In our approach, the inclusion of the antiferromagnetically ordered phase was done by resorting to 
the saddle-point analysis of the bosonic and fermionic effective actions, however
the general architecture of the method is not resting on this assumption. 
The method is suitable for general value of the correlation energy $U$, which is in
contrast to the TPSC method\cite{vilk1,vilk2} valid in a weak-coupling limit 
and slave boson (fermion) approach, where the normalization of the spectral
function is also violated.\cite{feng} 
In the large-$U$ limit, the theory is controlled by the parameter $J=4t^2/U$, in agreement
with previous approaches. 
Regarding the critical behavior, in the vicinity of the phase transition, our model based on the
spherical approach will be equivalent to that of $n$-vector model in the limit $n\rightarrow \infty$, 
which is the same as in the TPSC approach. 
We investigated the transformation properties of Hubbard Hamiltonian leading to a symmetry
adapted formulation, which explicitly exhibits the SU(2) and U(1)  invariance
included in the formalism.	
In this picture, the  collective bosonic modes that represent charge and spin  play an important role
since the are related to the underlying symmetries of the system and the ordered states.
The single-particle properties are obtained by
writing the original fermion field in terms of a U(1) phase field related to the charge, CP$^1$ bosons 
that parametrize the variable quantization axis related to the rotational symmetry and 
a pseudo-fermion.  This decomposition allows us to write down
the fermion Green's function by the product in real
space of the phase,  Schwinger boson propagator and the remaining fermionic propagator.
Because spatial correlations are now included, we find important
modifications of the electronic picture due to the formation of the ordered magnetic states.
We have shown that the single electron density of states.
consists of  two pieces. The first generates the 
the peak  which is a product of the condensate density of  CP$^1$ bosons that represents the antiferromagnetic order
and the and the pseudo-fermion spectral function. We found that this feature is an analog to the situation  present in 
the  normal state ARPES spectra in the underdoped samples, where the  the superconducting condensate 
produces similar behavior.

Finally, it would be interesting to extend our calculation to the more interesting doped system, 
which will require additional numerical effort. 

\acknowledgments
One of us (T.K.K) acknowledges the financial support 
from the  Ministry of Education and Science
MEN under grant No. 1PO3B 103 30 in the years 2006-2008.

\appendix

\section{Correlation Functions}

\subsection{Charge sector}

The Green's function in the charge sector reads:
\begin{equation}
G_{z}\left(\omega_{n}\right)=\frac{1}{\lambda_{z}+\frac{U}{8}
+\frac{2}{U}\left[\omega_{n}-i\frac{U}{2}f\left(\frac{2\mu}{U}\right)\right]^{2}}.
\end{equation}
Here, function $f\left(2\mu/U\right)=\mathrm{frac}\left(2\mu/U\right)-1,$
where $\mathrm{frac}\left(x\right)=x-\left[x\right]$ replaces summation
over winding number in Eq. (\ref{correlator}), which is valid for
temperatures $\beta\ll U$.
%
\subsection{Fermionic sector}
%
In the fermionic sector, the Nambu notation of the fermionic action
in Eq. (\ref{eq:FermionicSector_GreenFunction}) with vectors:
\begin{equation}
\bar{\Lambda}\left(\mathbf{k}\omega_{n}\right)
=\left[\bar{h}_{\uparrow\mathbf{k}},
\bar{h}_{\downarrow\mathbf{k}},
\bar{h}_{\uparrow\mathbf{k}-\bm{\pi}},
\bar{h}_{\downarrow\mathbf{k}-\bm{\pi}}\right]\left(\omega_{n}\right)
\end{equation}
leads to the Green's function matrix:
\begin{equation}
\mathbf{G}_{h\mathbf{k}}\left(\nu_{n}\right)
=\left[\begin{array}{cccc}
G_{h\mathbf{k}}^{\uparrow\uparrow} 
& 0 & F_{h\mathbf{k}}^{\uparrow\uparrow} & 0\\
0 & G_{h\mathbf{k}}^{\downarrow\downarrow} 
& 0 & F_{h\mathbf{k}}^{\downarrow\downarrow}\\
F_{h\mathbf{k}}^{\uparrow\uparrow} 
& 0 & G_{h\mathbf{k}-\bm{\pi}}^{\uparrow\uparrow} & 0\\
0 & F_{h\mathbf{k}}^{\downarrow\downarrow} 
& 0 & G_{h\mathbf{k}-\bm{\pi}}^{\downarrow\downarrow}
\end{array}\right]\left(\nu_{n}\right)
\end{equation}
with normal and anomalous Green's functions
 \begin{eqnarray}
G_{h}^{\alpha\alpha}\left(\mathbf{k}\nu_{n}\right)
 & = & \frac{\mu+2\chi\xi_{\mathbf{k}}-i\nu_{n}}
{\Delta_{c}^{2}+4\chi^{2}\xi_{\mathbf{k}}^{2}-\left(\mu-i\nu_{n}\right)^{2}},
\nonumber \\
F_{h}^{\alpha\alpha}\left(\mathbf{k}\nu_{n}\right)
 & = & \frac{\left(-1\right)^{\alpha}\Delta_{c}}
{\Delta_{c}^{2}+4\chi^{2}\xi_{\mathbf{k}}^{2}-\left(\mu-i\nu_{n}\right)^{2}}.
\end{eqnarray}
%

\subsection{Spin sector}
In the spin sector, the spin action in Eq. (\ref{eq:SpinSector_finalaction})
with vector
\begin{equation}
\bar{\Lambda}\left(\mathbf{k},\omega_{n}\right)=
\left[\bar{\zeta}_{\sigma\mathbf{k}},
\zeta_{\sigma-\mathbf{k}},\bar{\zeta}_{\sigma\mathbf{k}
-\bm{\pi}},\zeta_{\sigma-\mathbf{k}+\bm{\pi}}\right]\left(\omega_{n}\right)
\end{equation}
leads to spin Green's function matrix:
\begin{equation}
\mathbf{G}_{\zeta\mathbf{k}}\left(\omega_{n}\right)
=\left[\begin{array}{cccc}
G_{\zeta\mathbf{k}}, & F_{\zeta\mathbf{k}},
 & \tilde{G}_{\zeta\mathbf{k}}, & \tilde{F}_{\zeta\mathbf{k}}\\
F_{\zeta\mathbf{k}}, & G_{\zeta\mathbf{k}}, 
& \tilde{F}_{\zeta\mathbf{k}-\bm{\pi}}, & \tilde{G}_{\zeta\mathbf{k}-\bm{\pi}}\\
\tilde{G}_{\zeta\mathbf{k}}, & \tilde{F}_{\zeta\mathbf{k}-\bm{\pi}}, 
& G_{\zeta\mathbf{k}-\bm{\pi}}, & F_{\zeta\mathbf{k}-\bm{\pi}}\\
\tilde{F}_{\zeta\mathbf{k}}, & \tilde{G}_{\zeta\mathbf{k}-\bm{\pi}} 
& F_{\zeta\mathbf{k}-\bm{\pi}} & G_{\zeta\mathbf{k}
-\bm{\pi}}
\end{array}\right]\left(\omega_{n}\right),
\end{equation}
which elements read
\begin{eqnarray}
G_{\xi\mathbf{k}}\left(\omega_{n}\right) 
& = & \frac{1}{2}\left[\gamma_{-}\left(\mathbf{k},\omega_{n}\right)
+\gamma_{+}\left(\mathbf{k},\omega_{n}\right)\right]
\nonumber \\
\tilde{G}_{\xi\mathbf{k}}\left(\omega_{n}\right) 
& = & \frac{1}{2}\left[\gamma_{-}\left(\mathbf{k},\omega_{n}\right)
-\gamma_{+}\left(\mathbf{k},\omega_{n}\right)\right]
\nonumber \\
F_{\xi\mathbf{k}}\left(\omega_{n}\right) 
& = & -\frac{1}{2}\left[\gamma_{-}^{Q}\left(\mathbf{k},\omega_{n}\right)
+\gamma_{+}^{Q}\left(\mathbf{k},\omega_{n}\right)\right]
\nonumber \\
\tilde{F}_{\xi\mathbf{k}}\left(\omega_{n}\right)
 & = & \frac{1}{2}\left[\gamma_{-}^{Q}\left(\mathbf{k},\omega_{n}\right)
-\gamma_{+}^{Q}\left(\mathbf{k},\omega_{n}\right)\right],
\end{eqnarray}
where
\begin{eqnarray}
\gamma_{\pm}\left(\mathbf{k},\omega_{n}\right) 
& = & \frac{\frac{\omega_{n}^{2}}{\mathcal{E}_{s}}
\pm2i\theta\omega_{n}+\lambda_{\zeta}}{\left(\frac{\omega_{n}^{2}}{\mathcal{E}_{s}}
\pm2i\theta\omega_{n}+\lambda_{\zeta}\right)^{2}-4Q^{2}\xi_{\mathbf{k}}^{2}}
\nonumber \\
\gamma_{\pm}^{Q}\left(\mathbf{k},\omega_{n}\right)
 & = & \frac{2Q\xi_{\mathbf{k}}}{\left(\frac{\omega_{n}^{2}}{\mathcal{E}_{s}}
\pm2i\theta\omega_{n}+\lambda_{\zeta}\right)^{2}-4Q^{2}\xi_{\mathbf{k}}^{2}}.
\end{eqnarray}
%
\section{Spectral densities\label{sec:Appendix_Spectral-densities}}
%
The Green's function of the system is the combination of Green's functions
of charge, spin and fermionic sectors: 
\begin{eqnarray}
G_{c}^{\alpha\alpha'}\left(\mathbf{r}\tau,\mathbf{r}'\tau'\right) 
& = & \delta_{\alpha\alpha'}G_{z}\left(\mathbf{r}\tau,\mathbf{r}'\tau'\right)
 \nonumber \\
&\times &\left[G_{\zeta}^{11}\left(\mathbf{r}\tau,\mathbf{r}'\tau'\right)
G_{h}^{11}\left(\mathbf{r}\tau,\mathbf{r}'\tau'\right)\right.
\nonumber \\
 &+& \left.G_{\zeta}^{22}\left(\mathbf{r}\tau,\mathbf{r}'\tau'\right)
G_{h}^{22}\left(\mathbf{r}\tau,\mathbf{r}'\tau'\right)\right],
\end{eqnarray}
where its Fourier transform:
\begin{eqnarray}
G_{c\mathbf{k}}^{\alpha\alpha}\left(\omega_{n}\right) & = & \frac{1}{\beta N}\sum_{\mathbf{r}\ne\mathbf{r'}}\int_{0}^{\beta}d\tau e^{i\mathbf{k\left(r-r'\right)}+i\omega_{n}\left(\tau-\tau'\right)}\nonumber \\
 & \times & G^{\alpha\alpha}\left(\mathbf{r}-\mathbf{r'},\tau-\tau'\right).
\end{eqnarray}
 The spectral density is the imaginary part of the single-particle
Green's function and therefore contains full information
about the temporal and spatial evolution of a single electron
or a single hole in the interacting many-electron
system.
The spectral density is defined for fermions as follows
\begin{equation}
G_{X\mathbf{k}}^{\alpha\alpha}\left(\nu_{n}\right)
=\int_{-\infty}^{+\infty}\frac{d\omega}{2\pi}
\frac{A_{X\mathbf{k}}^{\alpha\alpha}
\left(\omega\right)}{i\nu_{n}-\omega}\label{eq:G_do_A_ferm}
\end{equation}
with $X=c,\, h$ for full system and fermionic sector, respectively.
Similarly, for bosonic sector:
\begin{equation}
G_{X\mathbf{k}}\left(\omega_{n}\right)=
\int_{-\infty}^{+\infty}\frac{d\omega}{2\pi}
\frac{A_{X\mathbf{k}}\left(\omega\right)}{i\omega_{n}-\omega},\label{eq:G_do_A_bos}
\end{equation}
where $X=z,\zeta$ for charge and spin part. The full spectral function
of the system expressed in Fourier variables is a double convolution
of three elementary spectral functions of charge, spin and fermionic
sectors, which written in terms of the real frequencies are
\begin{eqnarray}
A_{zh\mathbf{k}}^{\alpha\alpha}\left(\omega\right) 
& = & \frac{1}{N}\sum_{\mathbf{q}}\int_{-\infty}^{+\infty}\frac{d\omega'}{2\pi}A_{z\mathbf{q}}\left(\omega'\right)A_{h\mathbf{k}-\mathbf{q}}^{\alpha\alpha}\left(\omega-\omega'\right)\nonumber \\
 & \times & \left[n_{B}\left(-\omega'\right)+n_{F}\left(\omega-\omega'\right)\right]\nonumber \\
A_{zh\zeta\mathbf{k}}^{\alpha\alpha}\left(\omega\right) 
& = & \frac{1}{N}\sum_{\mathbf{q}\ne0}\int_{-\infty}^{+\infty}\frac{d\omega'}{2\pi}A_{\zeta\mathbf{q}}^{\alpha\alpha}\left(\omega'\right)A_{zh\mathbf{k}-\mathbf{q}}^{\alpha\alpha}\left(\omega-\omega'\right)\nonumber \\
 & \times & \left[n_{B}\left(-\omega'\right)+n_{F}\left(\omega-\omega'\right)\right].\label{eq:AzhiAzhz}
\end{eqnarray}
Introducing a density of states being a local ($\mathbf{k}$-integrated)
spectral density defined as
\begin{equation}
\rho_{c}^{\alpha\alpha}\left(\omega\right)=-\frac{1}{2\pi N}\sum_{\mathbf{k}}A_{c\mathbf{k}}^{\alpha\alpha}\left(\omega\right),
\end{equation}
one obtains convolution expressions
\begin{eqnarray}
\rho_{zh}^{\alpha\alpha}\left(\omega\right) & = & \int_{-\infty}^{+\infty}d\omega'\rho_{z}\left(\omega'\right)\rho_{h}^{\alpha\alpha}\left(\omega-\omega'\right)\nonumber \\
 & \times & \left[n_{B}\left(-\omega'\right)+n_{F}\left(\omega-\omega'\right)\right]\nonumber \\
\rho_{zh\zeta}^{\alpha\alpha}\left(\omega\right)
 & = & \int_{-\infty}^{+\infty}d\omega'\rho_{\zeta}^{\alpha\alpha}\left(\omega'\right)
\rho_{zh}^{\alpha\alpha}\left(\omega-\omega'\right)\nonumber \\
 & \times & \left[n_{B}\left(-\omega'\right)+n_{F}\left(\omega-\omega'\right)\right].
\end{eqnarray}
The density of states in the charge sector reads:
\begin{equation}
\rho_{z}\left(\omega\right)=\frac{\delta\left(\omega-E_{c}^{-}\right)-\delta\left(\omega-E_{c}^{+}\right)}{2\sqrt{f^{2}\left(\frac{2\mu}{U}\right)+\frac{2\delta\lambda_{z}}{U}}},
\end{equation}
while in the fermionic sector
\begin{eqnarray}
\rho_{h}^{\alpha\alpha}\left(\omega\right) & = & \frac{U^{2}\left|\omega-\bar{\mu}\right|}{4t^{4}v^{2}\left(B_{+}-B_{-}\right)}\Theta\left[\left(\omega-\bar{\mu}\right)^{2}-\Delta_{c}^{2}\right]\nonumber \\
 & \times & \left[\rho_{2D}\left(B_{-}\right)+\rho_{2D}\left(B_{+}\right)\right]\label{fermionic_density_states}
\end{eqnarray}
with
\begin{equation}
B_{\pm}=\pm\frac{U\sqrt{\left(\omega-\bar{\mu}\right)^{2}-\Delta_{c}^{2}}}{2t^{2}v}.
\end{equation}
In a non-dispersive case ($t=0$ or $v=0$), the fermionic density
of states
\begin{equation}
\rho_{h}^{\alpha\alpha}\left(\omega\right)=\frac{1}{2}\delta\left(\omega-\bar{\mu}+\Delta_{c}\right)+\frac{1}{2}\delta\left(\omega-\bar{\mu}-\Delta_{c}\right).
\end{equation}
In the spin sector, 
\begin{eqnarray}
 &  & \rho_{\zeta}\left(\omega\right)=\frac{1}{4Q}\left[\mathrm{sgn}\left(\theta-\frac{\omega}{\mathcal{E}_{s}}\right)\rho_{2D}\left(\frac{\frac{\omega^{2}}{\mathcal{E}_{s}}-\lambda_{\zeta}-2\theta\omega}{2Q}\right)\right.\nonumber \\
 &  & \,\,\,\,\,-\left.\mathrm{sgn}\left(\theta+\frac{\omega}{\mathcal{E}_{s}}\right)\rho_{2D}\left(\frac{\frac{\omega^{2}}{\mathcal{E}_{s}}-\lambda_{\zeta}+2\theta\omega}{2Q}\right)\right],
\end{eqnarray}
where
\begin{equation}
\rho_{2D}\left(x\right)=\frac{\Theta\left(1-\frac{x^{2}}{4}\right)}{\pi^{2}}\mathbf{K}\left(\sqrt{1-\frac{x}{4}^{2}}\right)
\end{equation}
is the density of states for the square lattice and $\mathbf{K}(x)$ stands for the complete
elliptic integral of the first kind.\cite{abram}
\section{Normalization condition}
%
Among the general properties of the spectral function
there are several sum rules. A fundamental one is 
\begin{equation}
\int_{-\infty}^{+\infty}\frac{d\omega}{2\pi}
A_{c\mathbf{k}}^{\alpha\alpha}\left(\omega\right)=1.\label{eq:Full_Norm_Simple}
\end{equation}
which reminds us that $ A_{c\mathbf{k}}^{\alpha\alpha}\left(\omega\right)$ describes the probability
of removing/adding an electron with momentum $\bf k$ and
energy $\omega$  to a many-body system.
Therefore, correctly calculated spectral function of the system should meet the
normalization condition. We can verify this is indeed the case  for the scheme presented in the present work.
Since, $A_{c\mathbf{k}}^{\alpha\sigma}\left(\omega\right)$ is given
by the Eq. (\ref{eq:SpectralDensity_Full}), its norm reads:
\begin{eqnarray}
 &  & \int_{-\infty}^{+\infty}\frac{d\omega}{2\pi}A_{c\mathbf{k}}^{\alpha\alpha}\left(\omega\right)\nonumber \\
 &  & =\int_{-\infty}^{+\infty}\frac{d\omega}{2\pi}
\left[m_0^{2}A_{zh\mathbf{k}}^{\alpha\alpha}
\left(\omega\right)+\sum_{\sigma}
A_{zh\zeta\mathbf{k}}^{\sigma\sigma}\left(\omega\right)\right].
\label{eq:Full_Norm}
\end{eqnarray}
Considering the first of two terms $A_{zh\mathbf{k}}^{\alpha\alpha}\left(\omega\right)$
and using the expression for the convolved charge-fermionic spectral
density from Eq. (\ref{eq:AzhiAzhz}), one obtains: 
\begin{eqnarray}
 &  & \int_{-\infty}^{+\infty}\frac{d\omega}{2\pi}A_{zh\mathbf{k}}^{\alpha\alpha}\left(\omega\right)=\frac{1}{N}\sum_{\mathbf{q}\ne0}\int_{-\infty}^{+\infty}\frac{d\omega}{2\pi}\frac{d\omega'}{2\pi}A_{z\mathbf{q}}\left(\omega'\right)\nonumber \\
 &  & \,\,\,\,\times A_{h\mathbf{k}-\mathbf{q}}^{\alpha\alpha}\left(\omega-\omega'\right)\left[n_{B}\left(-\omega'\right)+n_{F}\left(\omega-\omega'\right)\right].\end{eqnarray}
Transforming the integration variable $\omega\rightarrow\omega-\omega'$
we obtain:
\begin{widetext}
\begin{eqnarray}
\int_{-\infty}^{+\infty}\frac{d\omega}{2\pi}A_{zh\mathbf{k}}^{\alpha\alpha}\left(\omega\right) & = & \frac{1}{N}\sum_{\mathbf{q}\ne0}\left[\int_{-\infty}^{+\infty}\frac{d\omega'}{2\pi}A_{z\mathbf{q}}\left(\omega'\right)n_{B}\left(-\omega'\right)\right]
\left[\int_{-\infty}^{+\infty}\frac{d\omega}{2\pi}A_{h\mathbf{k}-\mathbf{q}}^{\alpha\alpha}\left(\omega\right)\right]\nonumber \\
 & + & \frac{1}{N}\sum_{\mathbf{q}\ne0}
\left[\int_{-\infty}^{+\infty}\frac{d\omega'}{2\pi}A_{z\mathbf{q}}\left(\omega'\right)\right]\left[\int_{-\infty}^{+\infty}\frac{d\omega}{2\pi}A_{h\mathbf{k}-\mathbf{q}}^{\alpha\alpha}\left(\omega\right)n_{F}\left(\omega\right)\right].
\end{eqnarray}
\end{widetext}
Since, the charge spectral function is antisymmetric
$A_{z\mathbf{q}}\left(\omega\right)=-A_{z\mathbf{q}}\left(-\omega\right)$,
its integral over frequencies vanishes: 
\begin{equation}
\int_{-\infty}^{+\infty}\frac{d\omega}{2\pi}A_{z\mathbf{q}}\left(\omega\right)=0.
\end{equation}
On the other hand, the integrand of fermionic spectral density is
normalized 
\begin{equation}
\int_{-\infty}^{+\infty}\frac{d\omega}{2\pi}A_{h\mathbf{k}-\mathbf{q}}^{\alpha\alpha}\left(\omega\right)=1.
\end{equation}
The remaining factor:
\begin{equation}
\frac{1}{N}\sum_{\mathbf{q}'\ne0}\int_{-\infty}^{+\infty}\frac{d\omega'}{2\pi}A_{z\mathbf{q}'}\left(\omega'\right)n_{B}\left(-\omega'\right)=1
\end{equation}
is simply a charge sector constraint, which can be checked by substituting
Eq. (\ref{eq:G_do_A_bos}) to Eq. (\ref{eq:Constraint_z}) and using
a summation rule:
\begin{equation}
\frac{1}{\beta}\sum_{n}\frac{e^{i0^{-}}}{i\omega_{n}-\omega}=\frac{1}{e^{-\beta\omega}-1}=n_{B}\left(-\omega\right).\end{equation}
Finally, the first term in the Eq. (\ref{eq:Full_Norm}) reads:
\begin{equation}
\int_{-\infty}^{+\infty}\frac{d\omega}{2\pi}A_{zh\mathbf{k}}^{\alpha\alpha}\left(\omega\right)=1.
\end{equation}
Similarly, one have to treat the second term in the Eq. (\ref{eq:Full_Norm})
\begin{eqnarray}
 &  & \int_{-\infty}^{+\infty}\frac{d\omega}{2\pi}A_{zh\zeta\mathbf{k}}^{\sigma\sigma}\left(\omega\right)=\frac{1}{N}\sum_{\mathbf{q}\ne0}
\int_{-\infty}^{+\infty}\frac{d\omega}{2\pi}\frac{d\omega'}{2\pi}A_{\zeta\mathbf{q}}^{\alpha\alpha}\left(\omega'\right)\nonumber \\
 &  & \,\,\,\,\times A_{zh\mathbf{k}-\mathbf{q}}^{\alpha\alpha}\left(\omega-\omega'\right)\left[n_{B}\left(-\omega'\right)+n_{F}\left(\omega-\omega'\right)\right].
\end{eqnarray}
Once again $A_{\zeta\mathbf{q}}^{\alpha\alpha}\left(\omega\right)
=-A_{\zeta\mathbf{q}}^{\alpha\alpha}\left(-\omega\right)$,
which leads to the vanishing of the integral
\begin{equation}
\int_{-\infty}^{+\infty}\frac{d\omega}{2\pi}A_{\zeta\mathbf{q}}\left(\omega\right)=0.
\end{equation}
The remaining part is the spin sector constraint
\begin{equation}
\frac{1}{N}\sum_{\sigma}\sum_{\mathbf{q}'\ne0}\int_{-\infty}^{+\infty}\frac{d\omega'}{2\pi}A_{\zeta\mathbf{q}'}^{\sigma\sigma}\left(\omega'\right)n_{B}\left(-\omega'\right)=1-m_0^{2}.
\end{equation}
It means that the second term in the Eq. (\ref{eq:Full_Norm}) is
equal to
\begin{equation}
\sum_{\sigma}\int_{-\infty}^{+\infty}\frac{d\omega}{2\pi}A_{zh\zeta\mathbf{k}}^{\sigma\sigma}\left(\omega\right)=1-m_0^{2}.
\end{equation}
Substituting the results to the Eq. (\ref{eq:Full_Norm}) one can
see that the normalization condition from the Eq. (\ref{eq:Full_Norm_Simple})
is always fulfilled. The norm of the full spectral density directly
depends on the constraints in the charge and spin bosonic sectors.
Therefore, careful solution of the self-consistent equations in Sec.
\ref{sec:Self-Consistent-Equations} is of primary importance to obtain
physically reasonable results.


\end{document}